\documentclass[a4paper,fleqn,usenatbib,longbibliography,amssymb]{mnras}
\usepackage{newtxtext,newtxmath}
\usepackage[T1]{fontenc}
\usepackage{ae,aecompl}

\usepackage{graphicx}	
\usepackage{amsmath}	
\usepackage{amssymb}	
\usepackage{mymacros}

\title[Twisted light from M87*]{Measurement of the spin of the M87 black hole from its observed twisted light}

\author[F. Tamburini et al.]{
Fabrizio Tamburini,$^{1}$\thanks{E-mail: fabrizio.tamburini@gmail.com}
Bo Thid\'e$^{2}$\thanks{E-mail: bt@irfu.se}
and Massimo Della Valle$^{3,4}$\thanks{E-mail: massimo.dellavalle@inaf.it}
\\
$^{1}$ ZKM, Lorenzstra{\ss}e 19, Karlsruhe, D-76135 Germany\\
$^{2}$ Swedish Institute of Space Physics, {\AA}ngstr\"om Laboratory, Box 537, SE-751 21 Uppsala, Sweden\\
$^{3}$ Capodimonte Astronomical Observatory, INAF-Napoli, Salita Moiariello 16, I-80131 Naples, Italy\\
$^{4}$ European Southern Observatory, Karl-Schwarzschild-Stra{\ss}e 2, D-85748 Garching bei M\"unchen, Germany.}

\date{Accepted XXX. Received YYY; in original form ZZZ}

\pubyear{2019}

\begin{document}
\label{firstpage}
\pagerange{\pageref{firstpage}--\pageref{lastpage}}
\maketitle

\setlength{\abovedisplayskip}{3pt}
\setlength{\belowdisplayskip}{3pt}
\setlength{\intextsep}{-1ex} 

\begin{abstract}
We present the first observational evidence that light propagating near a rotating black hole is twisted in phase and carries orbital angular momentum (OAM). This physical observable allows a direct measurement of the rotation of the black hole.
We extracted the OAM spectra from the radio intensity data collected by the Event Horizon Telescope from around the black hole M87* by using wavefront reconstruction and phase recovery techniques and from the visibility amplitude and phase maps.
This method is robust and complementary to black-hole shadow circularity analyses.
It shows that the M87* rotates clockwise with an estimated rotation parameter $a=0.90\pm0.05$ with $\sim 95\%$ confidence level (c.l.) and inclination $i=17^\circ \pm2^\circ$, equivalent to a magnetic arrested disk with inclination $i=163^\circ\pm2^\circ$. 
From our analysis we conclude, within a 6 $\sigma$ c.l., that the M87* is rotating.
\end{abstract}


\section{Introduction}
Most of the knowledge we have about the Universe comes from observing and
interpreting its electromagnetic (EM) radiation.  This ranges from visible
light caught by the naked eye, to radio and higher photon energy emissions
intercepted by advanced telescopes.  The information carried by the EM
field is encoded, naturally or artificially, into conserved quantities
(observables) of the field that are transported from the source to the
distant observer where the information can be decoded and recovered.
The standard observables are the well-known ten conserved quantities
that are concomitant with the ten-dimensional Poincar\'e group of Noether
invariants: the field energy (a single scalar), used, e.g., in radiometry,
linear momentum (three components of a vector), used in, e.g.,  radio
communications, angular momentum (three components of a pseudovector),
used only partially in radio science and technology, and the center of
energy (three components of a vector).

Of these observables, we used, for the first time,
the EM angular momentum, which is related to rotation and
torque action \citep{Torres&Torner:Book:2011}, to measure
the rotation of the M87* black hole in a novel way. This
observable has not yet been fully exploited in astronomy
\citep{Harwit:APJ:2003,Thide&al:Incollection:2011,Anzolin&al:AA:2008}.
The total angular momentum $\mathbfit{J}$ comprises two, in general
inseparable, quantities: $(1)~$the spin angular momentum $\mathbfit{S}$
(SAM), associated with photon helicity, i.e., the wave polarization, and
$(2)~ $the orbital angular momentum (OAM) $\mathbfit{L}$, associated with
the torque action and electromagnetic vorticity. EM waves carrying OAM
are also known as ``twisted waves''.  This property remains valid down
to the single photon level as superpositions of photon eigenstates,
each with a well-defined value of SAM and OAM, as described in
\citep{Molina-Terriza&al:NPHY:2007}. Only in the paraxial approximation
$\mathbfit{S}$ and $\mathbfit{L}$ behave as two separate variables.
OAM-carrying beams permit the encoding and extraction of more information
into and from the electromagnetic radiation emitted by a distant source
than any existing method that is based on the intensity of the field
only  \citep{Torres&Torner:Book:2011,Tamburini&al:NJP:2012}.  In 2011, a
theoretical-numerical study published by \citet{Tamburini&al:NPHY:2011},
showed that light and radio emitted from an accretion disk (AD)
around a massive rotating black hole, are expected to be endowed with
specific distributions of OAM due to the model of the AD and general
relativity (GR) effects such as Kerr spacetime dragging and mixing
and gravitational lensing.  An example is the Einstein ring, such as
that in the M87 galaxy as observed by the EHT team, which surrounds
the black hole. The Einstein ring is created by gravitationally
lensed beams that, experiencing an anamorphic transformation
\citep{Beckwith&Done:MNRAS:2005}, is also accompanied by a polarization
rotation \citep{Su&Mallett:APJ:1980} due to the gravitational Faraday
effect \citep{Dehnen:IJTP:1973}, image deformation, and rotation due
to the lensing of the curved spacetime as well as other modifications
in the phase wavefront, including gravitational Berry phase effects
\citep{Carini&al:PRD:1992,Feng&Lee:IJMPD:2001,Yang&Casals:PRD:2014}.

Light propagating near rotating black holes experiences a behaviour
that is analogous to that experienced in an inhomogeneous anisotropic
medium in a mechanism related to the Pancharatnam-Berry geometric
phase \citep{dezela}.
This involves spatially inhomogeneous
transformations of the polarization vector in the observation plane
of an asymptotic observer causing the lensed light to acquire OAM.
The OAM present in that light can be univocally characterized by the
so-called spiral spectrum \citep{Torner&al:OE:2005}, to be discussed
more in detail in the methods section, revealing the rotation of the BH.
Because of the properties of the Kerr metric, the OAM content in the EM
radiation lensed by the Kerr black hole (KBH) and the associated spatial
distribution of the azimuthal phase $\phi(\varphi)$, depend only on the
observables inclination $i$ of the KBH equatorial plane with respect to
the observer and the rotation parameter $a=J_\text{BH}/(m_\text{BH}c)$,
where $J_\text{BH}$ is the KBH angular momentum, but not on the mass
$m_\text{BH}$ of the BH \citep{Tamburini&al:NPHY:2011}.

Here we report the first observational confirmation of such GR-induced
OAM vorticity, detected in the public brightness temperature data
\citep{EHTdata} and also from the visibility amplitude and phase maps
from Event Horizon Telescope (EHT) observations at $\lambda=1.3$~mm
of the central compact radio source surrounding a supermassive black
hole M87* in the galaxy M87 of the Virgo Cluster released by EHT team
\citep{EHT1,EHT2,EHT3,EHT4,EHT5,EHT6}.  Exploiting this OAM we were able
to directly measure the spin, sense of rotation and inclination of M87*.

\section{Methods}
\label{sec:methods}
The key tool to determine the BH rotation parameter $a$ is the so called
spiral spectrum. \citet{Torner&al:OE:2005} ideated this method that
uniquely identifies the OAM content in the electromagnetic radiation,
using the fundamental physical property that any light beam can be
decomposed into a set of discrete orthogonal eigenmodes, each carrying
its own well-defined OAM quantity and known geometry.  A convenient
choice is Laguerre-Gaussian modes.  In cylindrical polar coordinates
$(\rh,\ph,z)$ these paraxially approximate modes describe an EM field
at $z$ with amplitude \citep{Barnett&Allen:OC:1994}
\begin{multline} \label{eq:LG}
 u_{m,p}(r, \theta, z) = \sqrt{\frac{2 p!}{\pi (p + m)!}}
 \frac{1}{w(z)} {\left[\frac{r \sqrt{2}}{w(z)}\right]}^l
 L_p^m \left[\frac{2 r^2}{w^2(z)}\right] 
 \\
 \exp {\left[\frac{-\mathrm{i} k r^{2}}{2R(z)}-\frac{r^2}{w(z)^2}\right]} \exp \left[-\mathrm{i} (2 p + m + 1) \arctan
 \left(\frac{z}{z_R}\right)\right] e^{-\mathrm{i} m \varphi },
\end{multline}
where $m$ is the azimuthal index that describes the $z$ component of
OAM, i.e. the number of twists in the helical wavefront (the topological
charge), $p$ is the radial node number of the mode, $z_R$ is the Rayleigh
range of the beam, $w(z)$ the beam waist, $R(z)$ the radius of curvature,
$L_p^m$ the associated Laguerre orthogonal polynomial, and $\varphi$ the
azimuthal phase of the beam.  The decomposition of a beam into orthogonal
modes will give the spectrum of OAM states carried by the EM wavefront, i.e.\
the spiral spectrum.

Thus, to estimate the rotation of M87*, 
this is, step by step, the procedure that we followed:
\begin{enumerate}
\item Using KERTAP software \citep{Chen&al:APJS:2015} we simulated
numerically different scenarios of an Einstein ring by varying the
emission parameters of the AD, the inclination of the BH spin $i$, the
BH rotation parameter $a$ of a KBH. For each simulation we obtained,
in the image plane of an asymptotic observer, the field intensity,
polarization and phase distributions and determine the content of
OAM from the spiral spectrum \citep{Torner&al:OE:2005}, relating the
physical properties of the BH system to that of the OAM distribution
\citep{Tamburini&al:NPHY:2011} with the asymmetry parameter $q$. This
quantity is the ratio of the height of the $m=1$ component with respect
to the $m=-1$ one found in each spiral spectrum. If $q>1$, then the
rotation is clockwise, whilst $q=1$ indicates no rotation and $q<1$
counterclockwise. This identifies the content of OAM in the field. From
the results of the numerical simulations, from each AD model chosen in
the simulations, we relate $q$ with $a$ fixed $i$ by using polynomial
interpolations.
\item From the EHT data we extract the intensity $I$ and reconstruct
numerically the spatial phase distribution $P$ of the field in the image
plane of an asymptotic observer by using well-known phase reconstruction
techniques based on the Transport of Intensity Equation (TIE)
\citep{Barbero:OENGR:2006,Schulze&al:OE:2012,Lubk&al:PRL:2013,Ruelas&al:JO:2018,Kelly:IJO:2018}
and build the spiral spectra.
\item We estimate the rotation parameter $a$ from the comparison of the
spiral spectrum of each numerical simulation with those obtained from EHT
data by using the asymmetry parameter $q$~\citep{Torres&Torner:Book:2011}.
\end{enumerate}

We can use this technique with OAM states to determine the rotation parameter $a$ building up the spiral spectrum for the M87* Einstein ring because 
the radio source has been spatially resolved at $1.3$~mm wavelength
by the $\sim10000$ km EHT radio-interferometric baseline. In this case, the wavefront is obviously not plane such as that emitted from a point-like source at infinity, even if the distance
from the source to the Earth is $D= 54.8\pm2.6\times10^6$ light years
($D\sim16.8$~Mpc). Hence, we can determine, within a good approximation,
the OAM content in the wavefront from the OAM spectrum of the
received radio waves. Furthermore, from the asymmetry
parameter $q$ of the observed OAM spectrum we can to determine the sense and
magnitude of the rotation of the M87* and its inclination, verifying the theoretical predictions made by
\citet{Tamburini&al:NPHY:2011}.

As seen from Eq. [\ref{eq:LG}], this spectrum of OAM states carried by the EM wavefront is obtained by
Fourier transforming the EM field with respect to the azimuthal angle
$\varphi$. Since our twisted-light/OAM method operates in angular momentum
space and uses the additional information encoded in the phase of the
OAM beam, it is complementary to methods based on analyses in configuration
($\varphi$) space, such as the analysis of the deviation from the circularity
of the M87* shadow adopted by EHT team
\citep{EHT1,EHT2,EHT3,EHT4,EHT5,EHT6,Bambi&al:ARX:2019},
a method that has so far not been able to yield any
conclusive results. In fact, the geometry of the shadow in Kerr metric \citep{Kerr:PRL:1963}
and the radius of the photon ring (the photon capture radius) changes with the ray orientation relative
to the angular momentum vectors with respect to that of a Schwarzschild BH.
This results in a deformation of the circular shape of the black hole's shadow
that was not possible to be determined with the available data. 
This deformation, even if small ($<4\%$), will be potentially detectable only with future EHT acquisitions \citep{EHT1}; the data released so far only allow EHT team to determine the inclination of the BH spin and the presence of a clockwise rotation, through the comparison of the experimental data with about 60,000 templates of the Einstein ring and BH shadow obtained from numerical simulations.

Moreover, our method makes use of the additional information encoded in the phase
of the OAM beam characterized by the asymmetry parameter $q$ 
of the spiral spectrum due to the vorticity induced by the Kerr metric and emitted by
a slightly larger region of the Einstein ring surrounding the BH, as explained more in detail
in  \citet{Tamburini&al:NPHY:2011} and in the Supplementary Material (SM). Let us follow our procedure.

(i) Numerical simulations of OAM from the Einstein ring of a KBH.
As initial step in the analysis and interpretation of the EHT data,
we build up, with the help of the freely available software package 
KERTAP \citep{Chen&al:APJS:2015}, a set of numerical simulations describing several scenarios of
black-hole accretion disks, as measured by an ideal asymptotic observer, 
to compare with the results of the analysis of the experimental data.
The value of the rotation parameter varied in the interval $0.5<a<0.99$ clockwise and counterclockwise and the inclination $1^\circ<i<179^\circ$. For the AD of the Einstein ring we adopted a thermalized emission with $0.1<\Gamma<2$ power spectra, chosen in agreement with ALMA observations
\citep{Doi&al:EPJWebofConferences:2013,EHT5} and a radial power law index that specifies the radial steepness of the profile, of $n_r=3$; we also simulated synchrotron emission, Compton scattering, and bremsstrahlung. 
Regarding the AD model used to fit with the observations, as discussed by the EHT team in Ref.~\cite{EHT6}, we also find that, in any case, the image properties are determined mainly by the spacetime geometry.
KERTAP provides a detailed description of the propagation of light in Kerr spacetime
\citep{Kerr:PRL:1963} with an error $\sim 10^{-7}$, in geometric units ($G=c=1$) and OAM routines $\sim10^{-12}$.

The OAM content is then numerically calculated from KERTAP output images and the Stokes polarization parameters $(U, Q, V)$, analyzing the parallel transport of the electric field and the Pancharatnam-Berry geometrical phase with the vectorial field technique developed by \citet{Zhang&al:SR:2015} where the radiation emitted from the simulated Einstein ring is a fully polarized vectorial vortex beam propagating in free space in the $z$ direction from the BH to the asymptotic observer.
The resulting spatial distribution of the azimuthal phase $\phi(\varphi)$ of the radiation used to calculate the OAM and the associated OAM spiral spectrum are invariant with respect to the mass $m_\text{BH}$ of the KBH and depend only on the parameters $i$ and $a$ \citep{Tamburini&al:NPHY:2011}. 

(ii) Analysis of EHT data with OAM techniques.
In order to construct the spiral spectrum \citep{Torner&al:OE:2005}, it is necessary to
    analyze, at each point in the image plane, the intensity (amplitude)
    and the phase and then calculate numerically the OAM content by
    interpolating the field with the different OAM modes described in  Eq. \ref{eq:LG}.
    OAM beams are detectable and can be accurately characterized with interferometric techniques 
    that directly measure amplitude, intensity and phase, 
as has been demonstrated experimentally in the radio domain
\citep{Thide&al:PRL:2007,Tamburini&al:APL:2011,Tamburini&al:NJP:2012}.

Since these kind of data are not all available from the EHT observations, in order to obtain the spatial phase distribution, we had to reconstruct the evolution of the wavefront with the well-known non-interferometric technique based on the TIE method and reduced error procedure \citep{Barbero:OENGR:2006,Schulze&al:OE:2012,Lubk&al:PRL:2013,Ruelas&al:JO:2018,Kelly:IJO:2018}.
The TIE equations (see Eq. 6 in SM par. 1.2) recovers the phase evolution between two or more consecutive intensity images of a stable source, in the paraxial approximation, when the source itself is spatially translated along the $z-$axis with respect to the observer. This is exactly what happens for when the images of M87* were taken, because of its relative motion with respect to the Earth (for more details see SM).

\begin{figure}
\centering
 \includegraphics*[width=.41\linewidth]{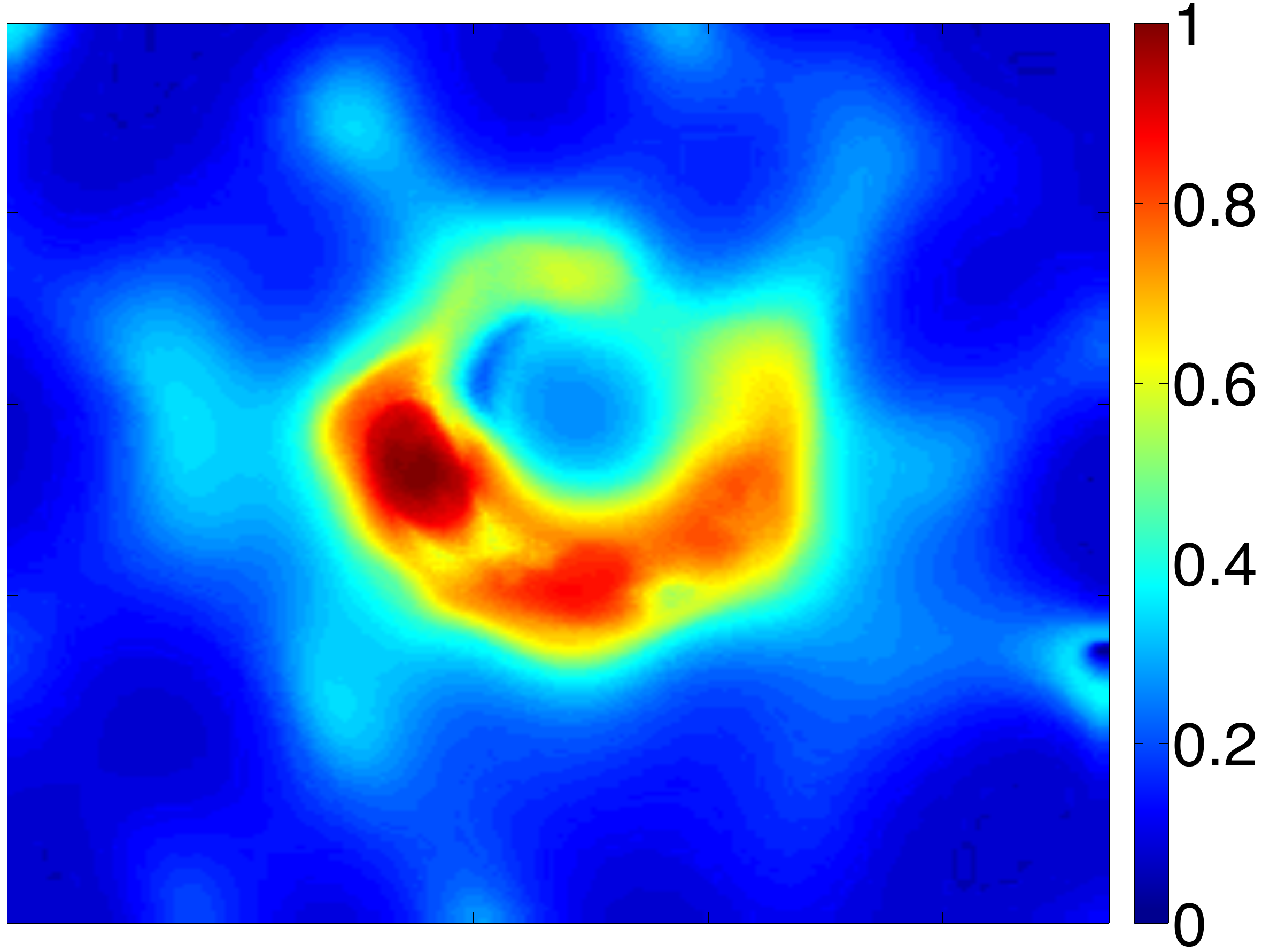}%
 \includegraphics*[width=.41\linewidth]{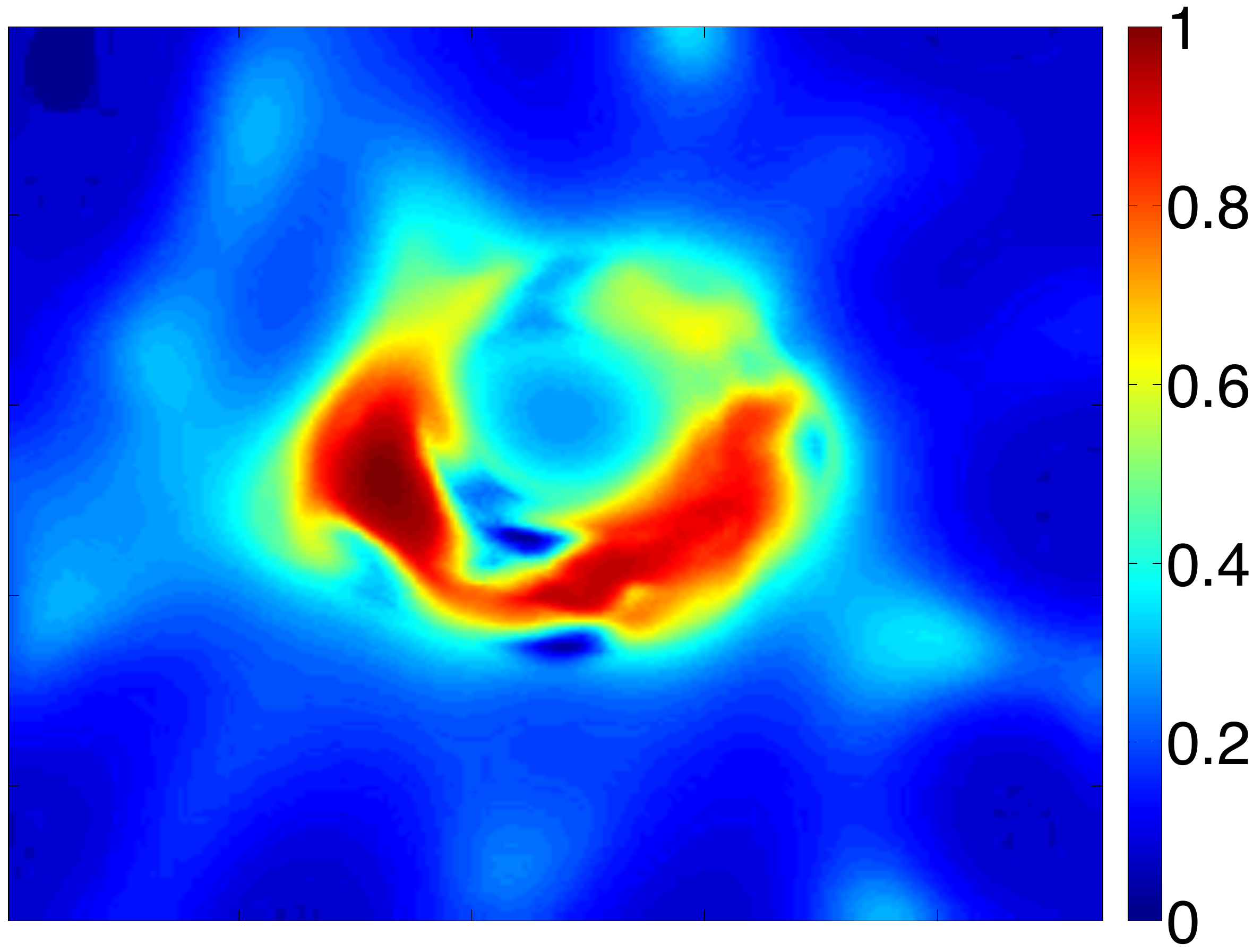}
 \includegraphics*[width=4cm,height=3.3cm]{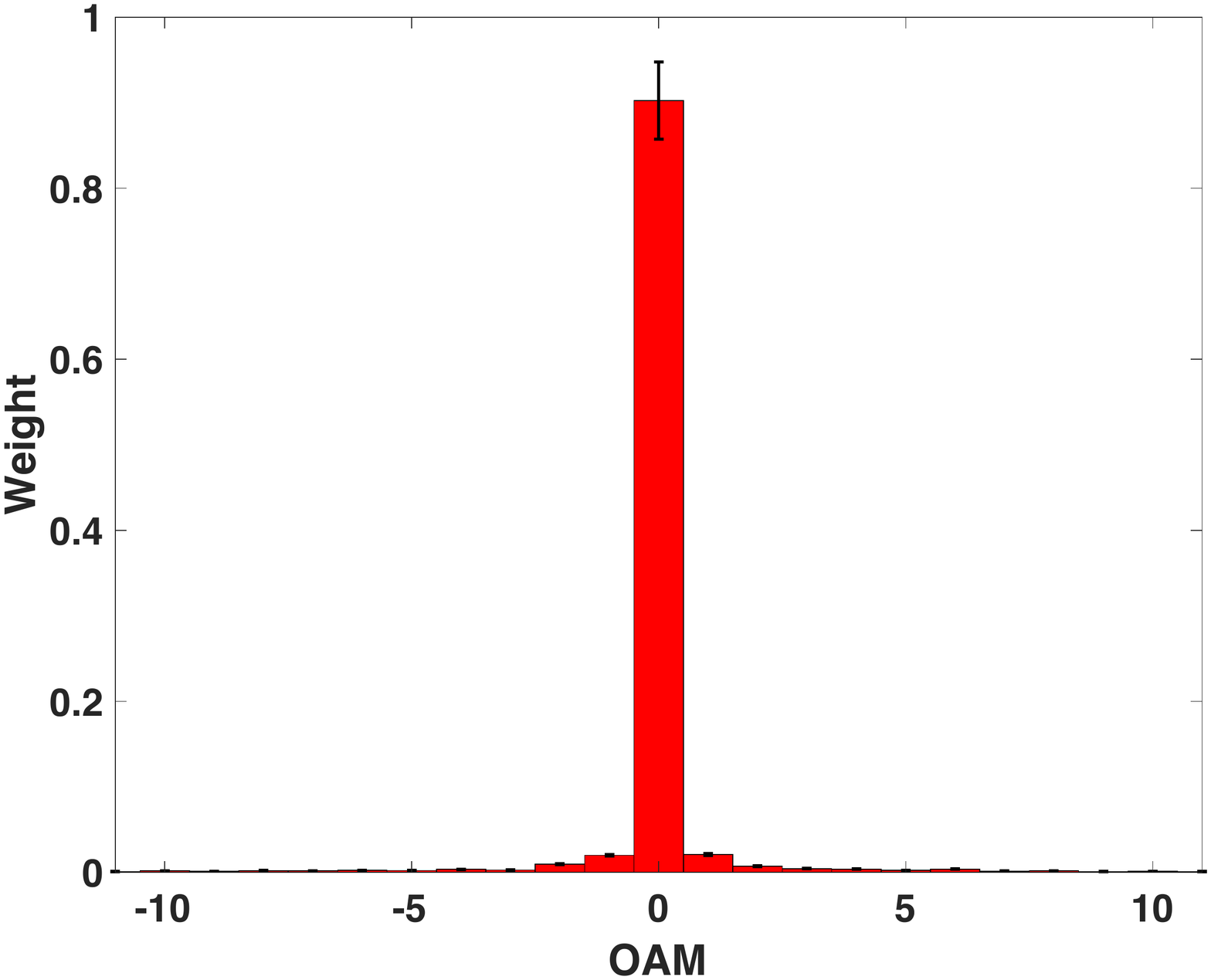}%
 \includegraphics*[width=4cm,height=3.31cm]{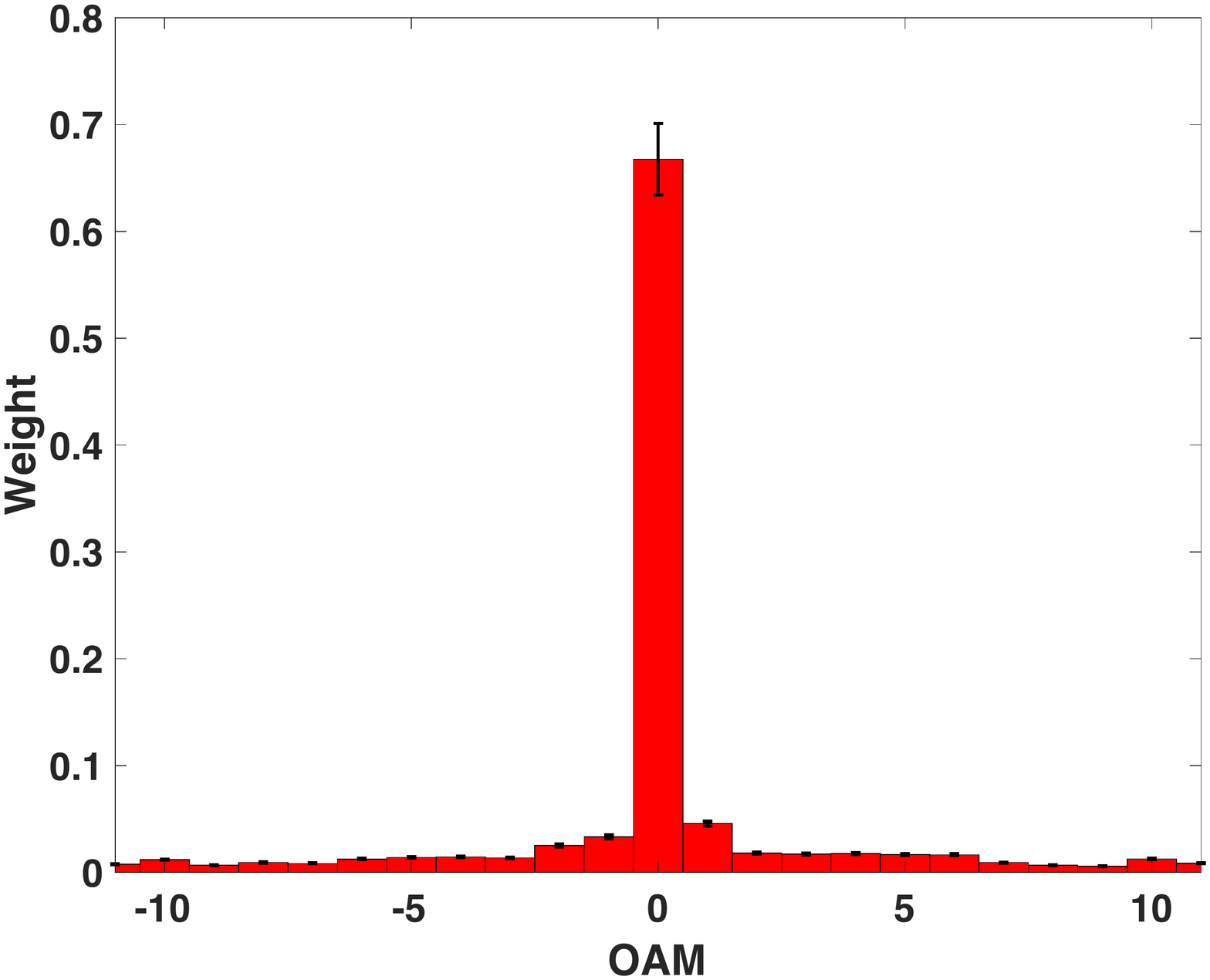}
 \caption{
 \textbf{Experimental results.}
  Normalized electric field component magnitudes along the observer's
  direction reconstructed from the TIE analysis of the brightness
  temperature in a finite frequency bandwidth and 
  for the corresponding spiral spectra for epoch~$1$ and epoch~$2$. 
  The asymmetry $q$ between the $m=1$ and $m=-1$ 
  components in both of the spiral spectra reveals the presence of
  twisted EM waves from the black hole Einstein ring \citep{Tamburini&al:NPHY:2011}
  with $a=0.904 \pm 0.046$ rotating clockwise. The spin is pointing away from
  Earth and an inclination between the approaching jet and the line of sight 
  of $i=17^\circ$ if the angular momentum of the accretion flow and that of 
  the black hole are anti-aligned (equivalent to a similar geometry with an inclination $i=163^\circ$,
  but where the angular momentum of the accretion flow and that of the
  BH are aligned). The Einstein ring has gravitational radius $r=5 R_g$, 
  as indicated by an EHT analysis dominated by incoherent emission \citep[see][and text]{EHT6,EHTdata}.
 The image coordinates are in arbitrary units.
 The intensity sidebars are normalized to unity.
 \label{fig:experiment}
 }
\end{figure}

The two  intensity and phase wavefront plots (see Fig. \ref{fig:experiment} in the main text and Fig. 1 and 2 in SM) have been reconstructed from two sets of two consecutive images 
of the Einstein ring of M87*. The two different observational runs of EHT are each separated by one
day, the first being epoch~$1$ (5 and 6 April 2017), and the second
being epoch~$2$ (10 and 11 April 2017). The data were obtained from
$10$ seconds of signal averaging as described in Refs.~\citep{EHTdata,EHT4,EHT5,EHT6}. 
The M87* exhibited a modest source evolution between the two pairs of nights
5--6 April and 10--11 April and a broad consistency within each pair.

Moreover, the TIE method can be applied because the physical evolution of the M87* Einstein ring structure during the two different EHT observation runs, each separated by one day, in two $2$~GHz frequency bands centered on $227.1$~GHz and $229.1$~GHz  is negligible and the source
can be considered stable as already stated in Refs.~\citep{EHT5,Doeleman&al:SCI:2012}.
More precisely, the evolution of the source during the time interval between the two sets
of pairs of days, was $<5\%$ within an observation \citep{EHT3,EHT4}.
Relative to the object, each day is $2.8R_g{}c^{-1}$ long for a BH
with mass $M=6.5\pm0.2|_\text{stat}\pm0.7|_\text{sys}\times10^9~M_\odot$.
This timescale is shorter than the crossing time of light of the
source plasma and short compared to the decorrelation timescale of EHT
simulations used to analyze the experimental results,
$50R_g{}c^{-1}$  \citep{EHT5}.
It was observed only a small evolution with a maximum of a few percent of difference between the two
epochs, separated by five days \citep{EHT6,EHTdata}.

With the TIE method one can estimate the OAM content from the brightness temperature distributions reported in the images \citep{Rybicki&Lightman:Book:RadiativeProcesses:2004} and then the rotation parameter of the KBH as it were obtained from two different and independent acquisitions.
The TIE method remains valid even though the shift $s$ along $z$ due to the motion of our planet around the Sun in the interval of one day appears to be enormous with respect to the wavelength
because the phase profile of an OAM wave repeats in space with a recurrence modulo $\lambda$.
Moreover, the properties of OAM waves are not degraded when traveling in free space the distance $s$, which is only a small fraction of the distance from M87* to the Earth.

The correctness of our procedure that uses the relative translational motion of 
the Earth with respect to M87* is mathematically ensured by the Wold theorem,
\citep{Wold:Book:StudyAnalysis:1954,Anderson:Book:StatisticalAnalysis:1994} 
as explained in the SM. 

(iii) Estimating the BH rotation parameter.
The rotation parameter $a$ can be estimated by comparing the asymmetry 
parameter $q$, given by the ratio of the OAM spectrum components $m=1$ and $m=-1$,
for any value of the inclination parameter $1^\circ< i <179^\circ$ with step of $1^\circ$ 
obtained from the numerical simulations 
and the values of $q$ obtained from the analysis of EHT data of epoch 1 and 2.
In Tab. \ref{tab:asym_ratio} is reported an example of values of the parameter $q$ obtained
from the numerical simulations of the Einstein ring of a KBH for two values of the 
inclination parameter, $i=17^\circ$, as indicated by EHT, and $i=46^\circ$ obtained by varying the rotation parameter $a$. The variation of $i=1^\circ$ can be adopted as indetermination in $i$ as it induces a small change $\Delta q= 0.0012$ corresponding to $\Delta a \simeq 0.06$ compatible with the variation of $a$ in its error interval.
The spiral spectra reported in Fig. \ref{fig:experiment} obtained with the TIE 
imaging techniques reveal the presence of an asymmetric ring with clockwise
rotation and a ``crescent'' geometric structure that exhibits a clear
central brightness depression.  This indicates a source dominated by
lensed emission surrounding the black hole~shadow.

\begin{table}
\caption{Table of values of the asymmetry parameter $q$, obtained by dividing the 
height of the $m=1$ by that of $m=-1$ components of the spiral spectra (see text and SM)
of Kerr black holes neighborhoods with different rotation parameters $0.5<a<0.9$ Vs. the rotation parameter $a$ for two sample inclinations $i=17^{\circ}$ and $i=46^{\circ}$ simulated with KERTAP. 
The rotation is clockwise and averaged over different accretion disk emission mechanisms characterized by $0.1<\Gamma<2$ power spectra. The error is $\sim10^{-7}$. In agreement with EHT we choose $i=17^\circ$ (and $i=163^\circ$).}
\label{tab:asym_ratio}
	\centering
	\begin{tabular}{l c c c r}
	\hline
 BH rotation\\  parameter $a$ & 0.50 & 0.60 & 0.80 & 0.90\\
\hline
 Asymmetry parameter $q$\\
 $i={17}^{\circ}$ &1.295 & 1.320 & 1.356 & 1.391\\
 $i={46}^{\circ}$ &1.392 & 1.412 & 1.424 & 1.438\\
 \hline
\end{tabular}
\end{table}

From the analysis of the two data sets we obtain the asymmetry
parameters $q_1=1.417 \pm 0.049$ for epoch~$1$ and $q_2=1.369 \pm 0.047$ for
epoch~$2$. They yield an averaged asymmetry in the spiral spectrum of
$\overline{q}=1.393\pm0.024$ corresponding to $a=0.904 \pm 0.046$ (see SM). 
The average value of the asymmetry parameter is compatible 
with that resulting from our numerical simulations, 
$q_\text{numerical}=1.389$, for an Einstein ring with
a radius of $5$ gravitational radii, emitting partially incoherent
light around a Kerr black hole.

The values of $q$ indicate that the KBH can have an inclination of 
$i=17^\circ \pm 2^\circ$, with the angular momenta of
the accretion flow and of the black hole anti-aligned, showing clockwise
rotation as described in Ref.~\citep{EHT5}, where an error of $1^\circ$ was reported.
The preferred inclination mentioned above is alternatively equivalent to a magnetically arrested disk (MAD) scenario with an inclination $i=163^\circ  \pm 2^\circ$ and the angular momentum of the accretion disk flow instead aligns with that of
the black hole \citep{Sobyanin:MNRASL:2018}. This is in agreement with
the results of the EHT collaboration, presented and discussed in
Refs.~\citep{EHT1,EHT2,EHT3,EHT4,EHT5,EHT6}.
EHT analysis suggests that the X-ray luminosity
is $\left\langle L_\text{X}10^{-2\sigma} \right\rangle <4.4\times10^{40}$~erg/s and the jet
power is $P_\text{jet}>10^{42}$~erg/s. The radiative efficiency is
smaller than the corresponding thin disk efficiency \citep{EHT5}.

\begin{figure}
\centering
 \includegraphics*[width=4cm,height=3.1cm]{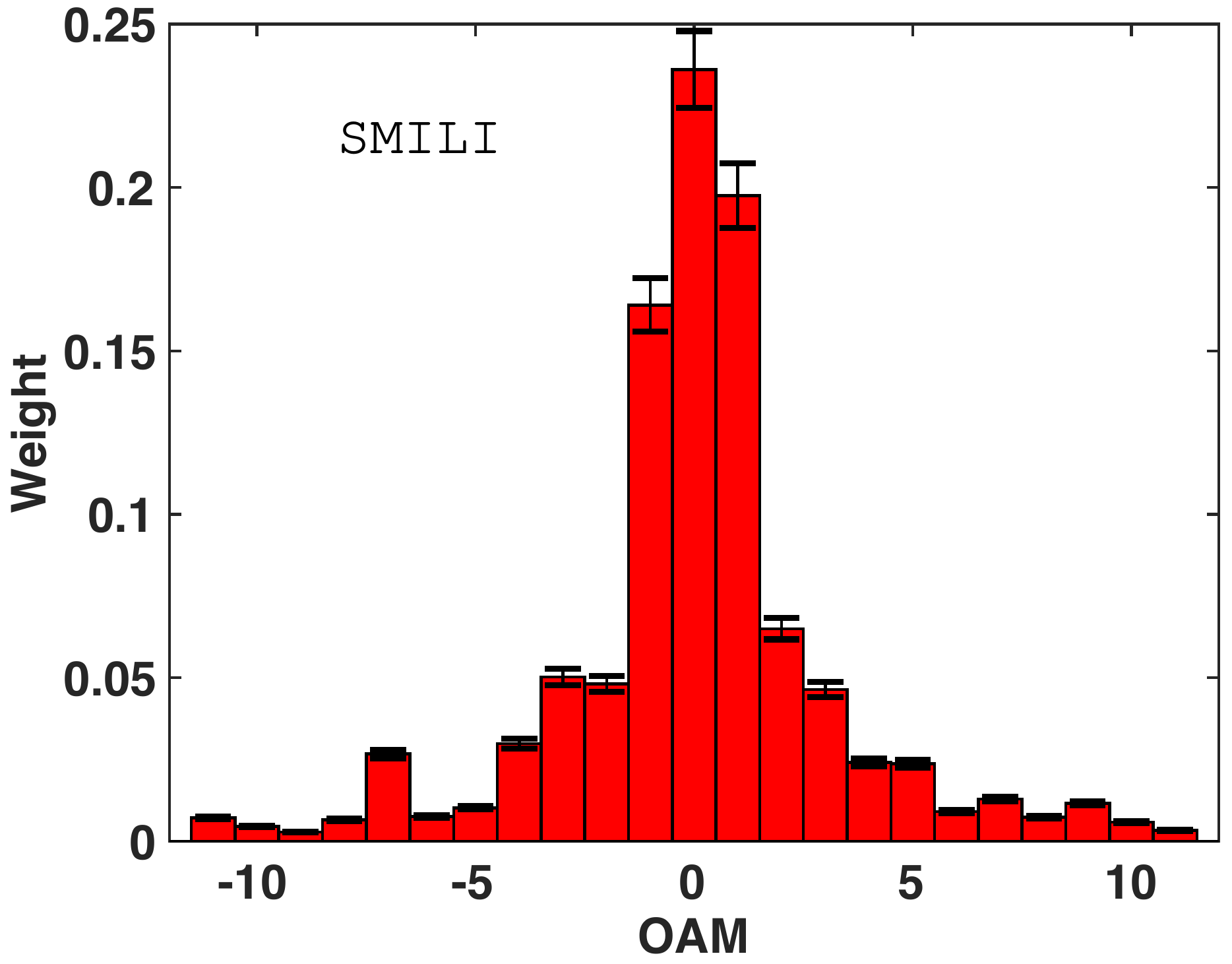}\hfill%
 \includegraphics*[width=4cm,height=3.1cm]{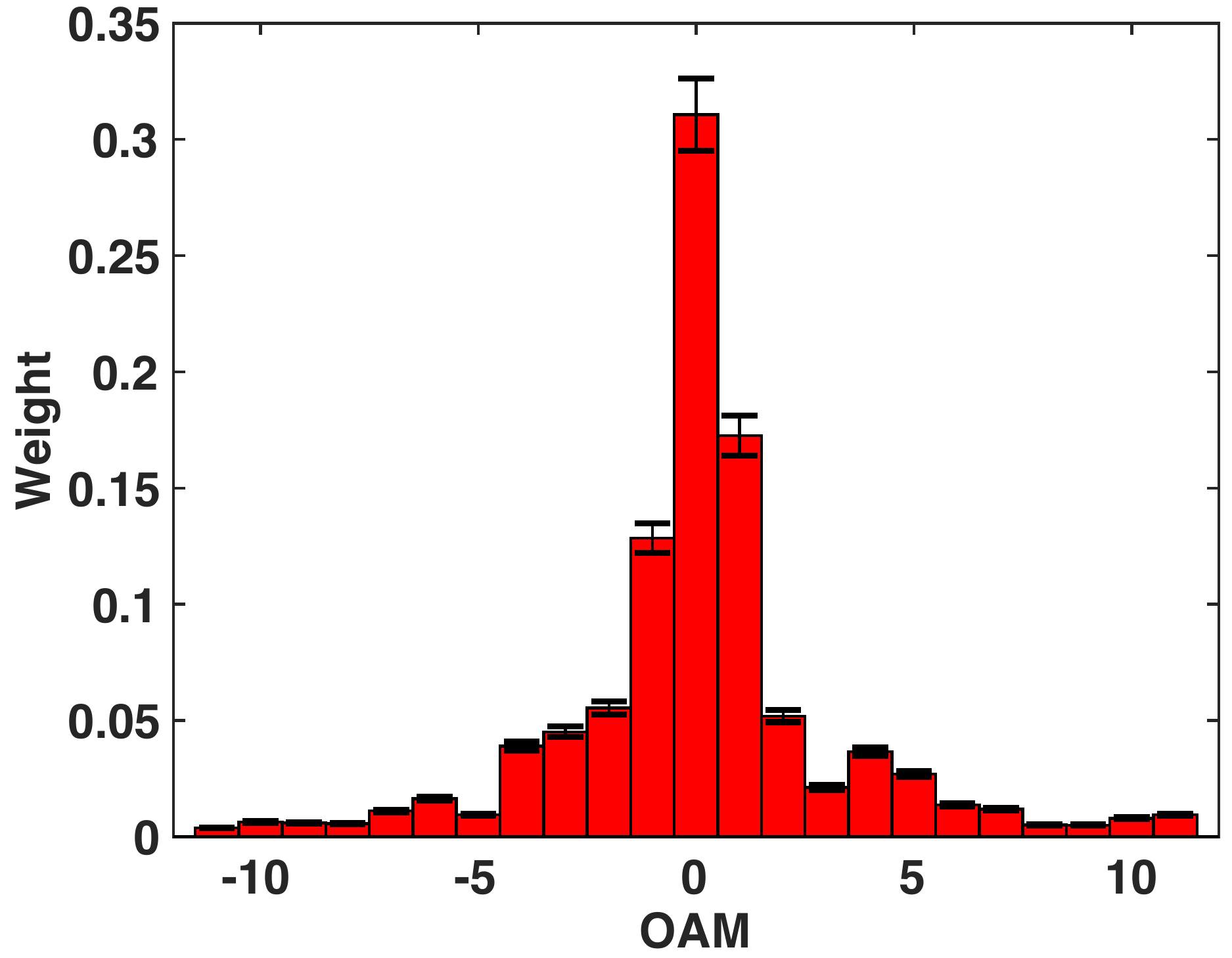}\\[2ex]
 \includegraphics*[width=4cm,height=3.1cm]{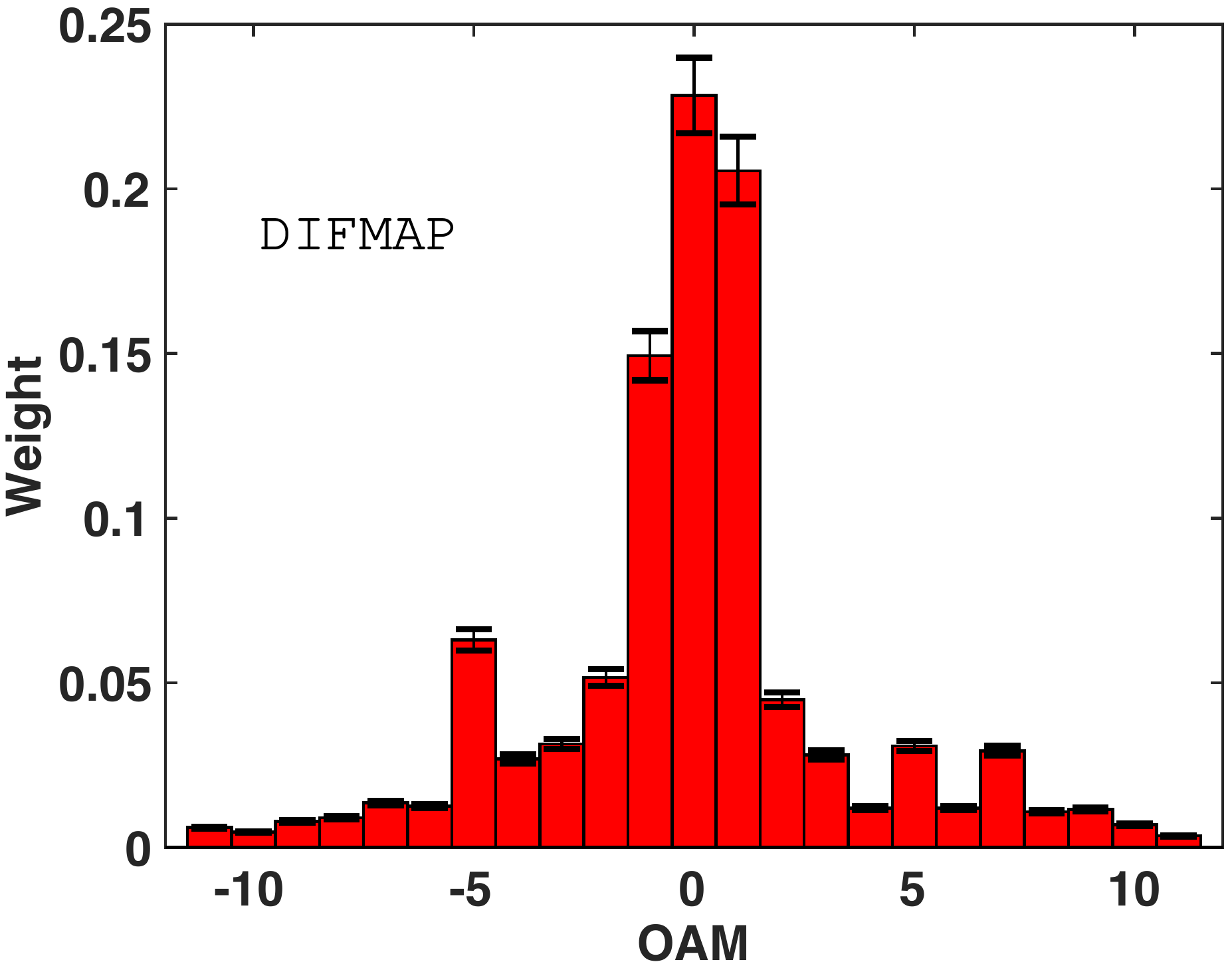}\hfill%
 \includegraphics*[width=4cm,height=3.1cm]{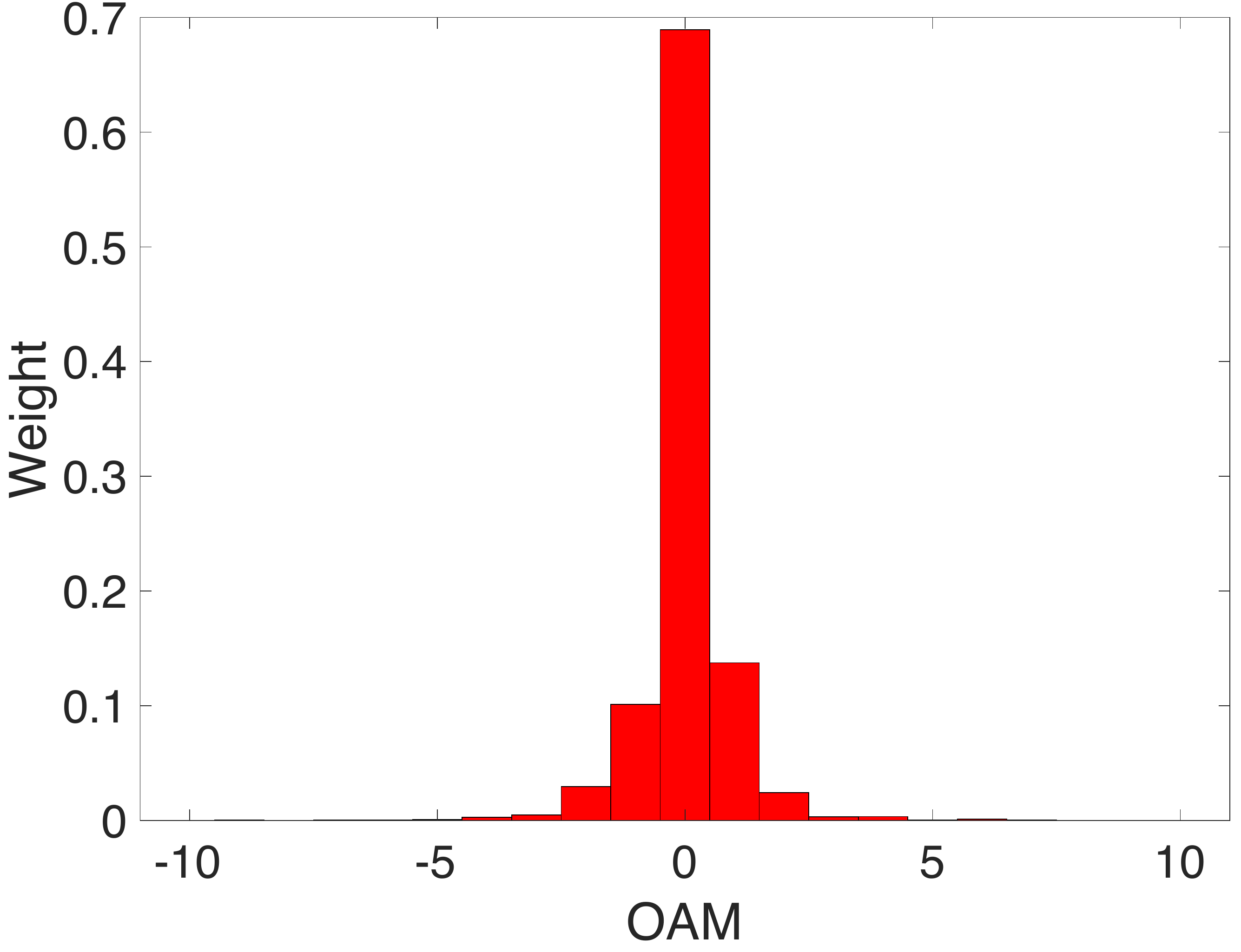}%
 \caption{
  OAM spectra from the results of SMILI, EHT (up) and DIFMAP (bottom left) 
  data analysys by EHT of the observation taken in 11 April 2017 \citep{EHT4,EHT6,EHTdata}
  and from the numerical simulations with KERTAP.
  The data considered here represent the visibility amplitude and phase, provided by EHT,
  as a function of the vector baseline. As the spatial phase distribution is, in this case,
  present in the data, we do not need to apply the TIE method to recover it. 
  In all the data sets the asymmetry parameter, the
  ratio $q$ between the $m=1$ and $m=-1$ peaks in the spiral spectra,
  is $q>1$ indicating clockwise rotation with $i=17^\circ$ if the angular
  momentum of the accretion flow and that of the black hole are anti-aligned
  (equivalent to a similar geometry with $i=163^\circ$,
  with angular momentum of the accretion flow and that of the
  BH are aligned). Bottom right, the value of $q$, obtained 
  from the spiral spectrum of KERTAP simulations, has an error of $\sim 10^{-7}$ 
  in agreement with the values found from these results and the TIE method.}
\label{fig:data}
\end{figure}

Our results above, obtained using the TIE methodology, show good agreement with those resulting from a more constraining approach, i.e. that which directly utilises the visibility amplitude and the phase maps for the day 11th April 2017 released by EHT. In particular, the ETH collaboration applied three data analysis methods: DIFMAP, EHT and SMILI, \citep{EHT4,EHT6,EHTdata}. 
We obtained $q_{\text{DIFMAP}}=1.401\pm 0.047$, $q_{\text{EHT}}=1.361\pm0.046$ and $q_{\text{SMILI}} = 1.319\pm 0.045$, yielding for this day an averaged value of $\overline{q}=1.360\pm0.027~(a=0.821\pm 0.062)$.
This value deviates by a quantity $\Delta a \simeq 0.087$ from that of epoch~$2$ obtained with
the TIE method. Since, in all cases, it results q > 1, that confirms the
presence of a clockwise rotation.
The spiral spectra of this additional data analysis are reported in Figure~\ref{fig:data}.

The error in the final estimate of $a$ depends on the numerical error of the 
TIE reconstruction method, $\left|\Delta{a}\right|_\text{TIE}\sim 10^{-7}$, 
and from that of the final EHT data products that include the calibrated total intensity
amplitude and phase information.
The latter are validated through a series of quality
assurance tests and show consistency across pipelines setting limits on
baseline systematic errors of $2\%$ in amplitude and $1^\circ$ in phase
giving a maximum error chi-squared test of $\sim 5\%$ \citep{EHT1,EHT2,EHT3,EHT4}.
By averaging  for any point in the image the values in a neighborhood of three pixels
we reduce the error in the calculation of the OAM spectrum $\left|\Delta{a}\right|_\text{EHT} \simeq 0.046$.
This ensures a good precision in the reconstruction of the surroundings of the BH and for our OAM analysis.
The total error of
 $a$ becomes,
\begin{equation}
\left|\Delta{a}\right|_\text{tot}  =\left|\Delta{a}\right|_\text{TIE} +\left|\Delta{a}\right|_\text{EHT} \simeq 0.046 ~,
\end{equation}
hence yielding a first conservative estimate of the rotation parameter at 
$a=0.904\pm0.046$, with $\sim 95\%$ c.l. and the hypothesis of a static BH is excluded at $\sim 6 \sigma$. 
More details can be found in the SM.

Even if the spatial phase profile was not measured directly for the values of $q$ obtained with the TIE
method, something
that can be achieved with interferometric or other more direct
techniques, the values of the rotation parameter $a$ obtained with
the OAM method agrees with the experimental data obtained from results
presented in the literature. Preferably, one should take a succession
of snapshots with the EHT at much shorter time intervals, typically a
few minutes rather than days, to improve the accuracy of the terms in
the TIE of Eq. 6 in the SM. This could be realized quite
straightforwardly.

With dedicated observations
and new fiducial pipeline images of amplitude and phase plots, 
and new interferometric methods that analyze the EM field
in intensity and phase 
(readily realized with standard radio telescopes
as was reported for the fiducial pipeline images of the visibility amplitude and phase
plots for 11 April 2017),
one should be able to drastically improve the
measurements of the OAM content within the spiral spectrum of  
the EM radiation emitted in the neighborhood of rotating black holes. 
As second step, it will be possible to estimate in a more accurate way
the BH rotation and the other fundamental parameters of the system 
from the shape of the shadow.  
Ultimately, radio telescopes equipped with antenna systems
optimized to directly capture and resolve the EM angular momentum in
the signals received should be able to unleash the full potential of
the OAM in observational astronomy.
One advantage of involving OAM in black-hole astronomy is that
polarization and OAM together build up the total angular momentum
invariant $\mathbfit{J}$ and when polarization is affected by the presence of
polarizing media such as dust and unstructured plasma, OAM will be less
affected and can be used as a reference point to extract additional
information about the source from its emitted light.

\section{Conclusions.}
Exploiting the properties of the Kerr metric, we were able to measure 
the rotation of M87* by analysing a sequence of images acquired 
by the Event Horizon Telescope. In particular, the estimate of the
rotation parameter $a$ was derived from the characterization of
the vorticity of the electromagnetic waves emitted from the surroundings
of the BH, affected by a strong gravitational lensing from the rotating
compact object \citep{Tamburini&al:NPHY:2011}.
The OAM and the electromagnetic vorticity were reconstructed from the public released
images of the brightness temperature of the Einstein ring of the M87 black
hole, taken during four different days with a well-known technique based on
the Transport of Intensity Equation 
\citep{Rybicki&Lightman:Book:RadiativeProcesses:2004,Lubk&al:PRL:2013,Zhang&al:SR:2015}
that permits the reconstruction of the phase wavefront and of the orbital
angular momentum content from two or more consecutive acquisitions of
the same source taken at different distances. 

By applying the general relativistic effect in the Kerr metric discussed in \citep{Tamburini&al:NPHY:2011}, we find that the central black hole in M87 is rapidly rotating clockwise with rotation parameter $a=0.90\pm0.10$  and inclination $i=17^\circ\pm 2^\circ$ (equivalent to a magnetic arrested disk with inclination $i=163^\circ\pm 2^\circ$) with $95\%$ c.l. and the case for a non-rotating BH is excluded with $\sim 6 \sigma$ c.l.. 
This enormous amount of energy trapped in the black hole rotation
implies a rotational energy of about $10^{64}$ erg, comparable to the
energy emitted by the brightest quasars over Gyr timescales \citep{Christodoulou&Ruffini:PRD:1971}.

\section*{Acknowledgements}
We thank Guido Chincarini for help and useful discussions.
FT. acknowledges ZKM and Peter Weibel for the financial
support. The encouragement and support from Erik~B.\,Karlsson is
gratefully acknowledged.

\vspace{100pt}

\textbf{{\LARGE Supplementary Material: }}

\section{OAM/phase reconstruction methods from numerical simulations
and experimental data}

The reconstruction of the phase profile from phase-front intensity
patterns is a novel non-interferometric technique. It is based on
the Transport of Intensity Equation (TIE) method that requires
a careful mathematical and numerical analysis of the problem
\citep{Lubk&al:PRL:2013,Ruelas&al:JO:2018}.  In our case we limit our
analysis to paraxial optics.  In principle, the TIE method requires
only two or more spatial intensity distributions (images) from a field
intensity acquisition system, exactly as was made by the Event Horizon
Telescope (EHT) collaboration.

In fact, the EHT collaboration took a series of radio ``snapshots'' at
different times. Because of the relative motion in space of the Earth and
M87*, as if the whole acquisition system were mounted over an movable
optical translation stage, the different observations were made at
different distances from the source.  This relative motion gave
rise to a translation in space that is enormous in terms of wavelengths
($\lambda=0.0013$ metres, $230$~GHz). Even so, the TIE method can still
be applied thanks to the following:
\begin{itemize}

 \item
 The source was stable during the acquisitions that were separated by one
 day in each of the two different observation epochs.

 \item
 The scattering due to interstellar matter did not affect the propagation
 of the radio waves from M87* to the Earth.

 \item
 The optimization procedure adopted for the image reconstruction minimised
 acquisition errors during the observations.

 \item
 The Einstein ring observed around M87* had a simple structure.
\end{itemize}
These furtuitous conditions ensure that the results of the excellent image
reconstruction process used by EHT team are good enough to allow a
reliable determination of the spatial phase distribution, enabling the
estimation of the black hole (BH) rotation parameter from an analysis
of the spiral spectrum components.

\subsection{Reconstruction of the spatial phase distribution from Stokes
and Pancharatnam--Berry phases}

The KERTAP software package \citep{Chen&al:APJS:2015} describes the
spacetime geometry around a rotating Kerr black hole \citep{Kerr:PRL:1963}
and calculates the gravitational optics phenomena such as lensing and
polarisation.  Here we describe how to use KERTAP to calculate the radio
orbital angular momentum induced by the Kerr black hole rotation.

We calculate the expected phase pattern of an Einstein ring for different
values of the rotation parameter, i.e.\ the angular momentum per unit mass
($a\leq1$), of the BH that, together with the BH mass $m_\text{BH}$,
is described in Boyer--Lindquist coordinates in geometric units $(G=c=1)$ 
by the Kerr space-time in the cylindrical coordinate system $(t,r,\th,\ph)$
\begin{multline}
 \ds^2 =
 \frac{\varrho}{\Delta}\dr^2
 +\varrho^2\d\th^2       
 +\frac{\sin^2\th}{\varrho^2}\left[a\dt-\left(r^2+a^2\right)\d\ph\right]^2
\\
 -\frac{\Delta}{\varrho^2}\left(\dt-a\sin^2\th\d\ph\right)^2
\end{multline}
where
\begin{align}
 \varrho^2 = r^2 + a^2\cos^2\th
\end{align}
and
\begin{align}
 \Delta = r^2 -2m_\text{BH}r+a^2
\end{align}

\begin{figure*}
\centering
 \hfil
\includegraphics*[width=.5\linewidth]{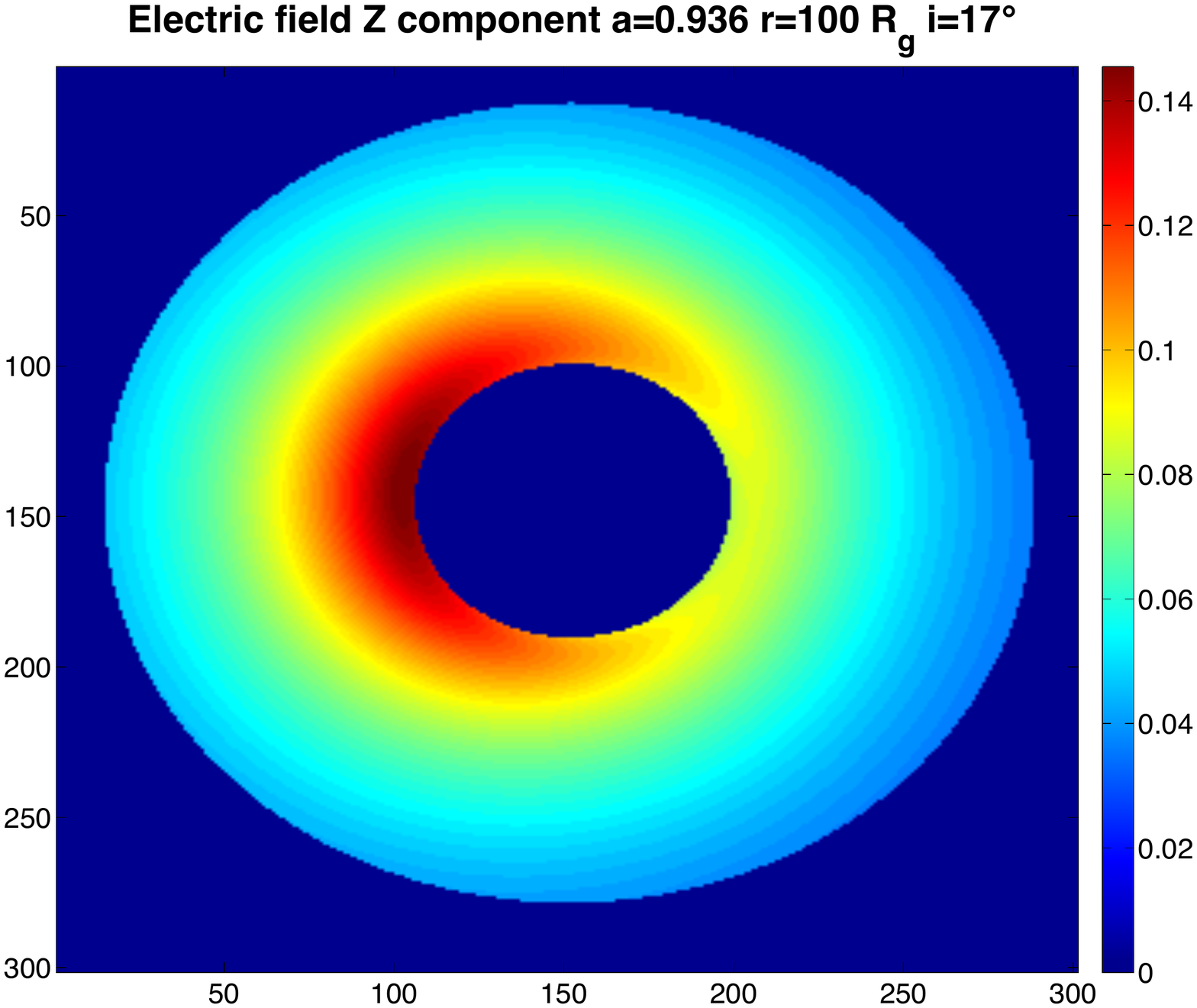}%
\includegraphics*[width=.5\linewidth]{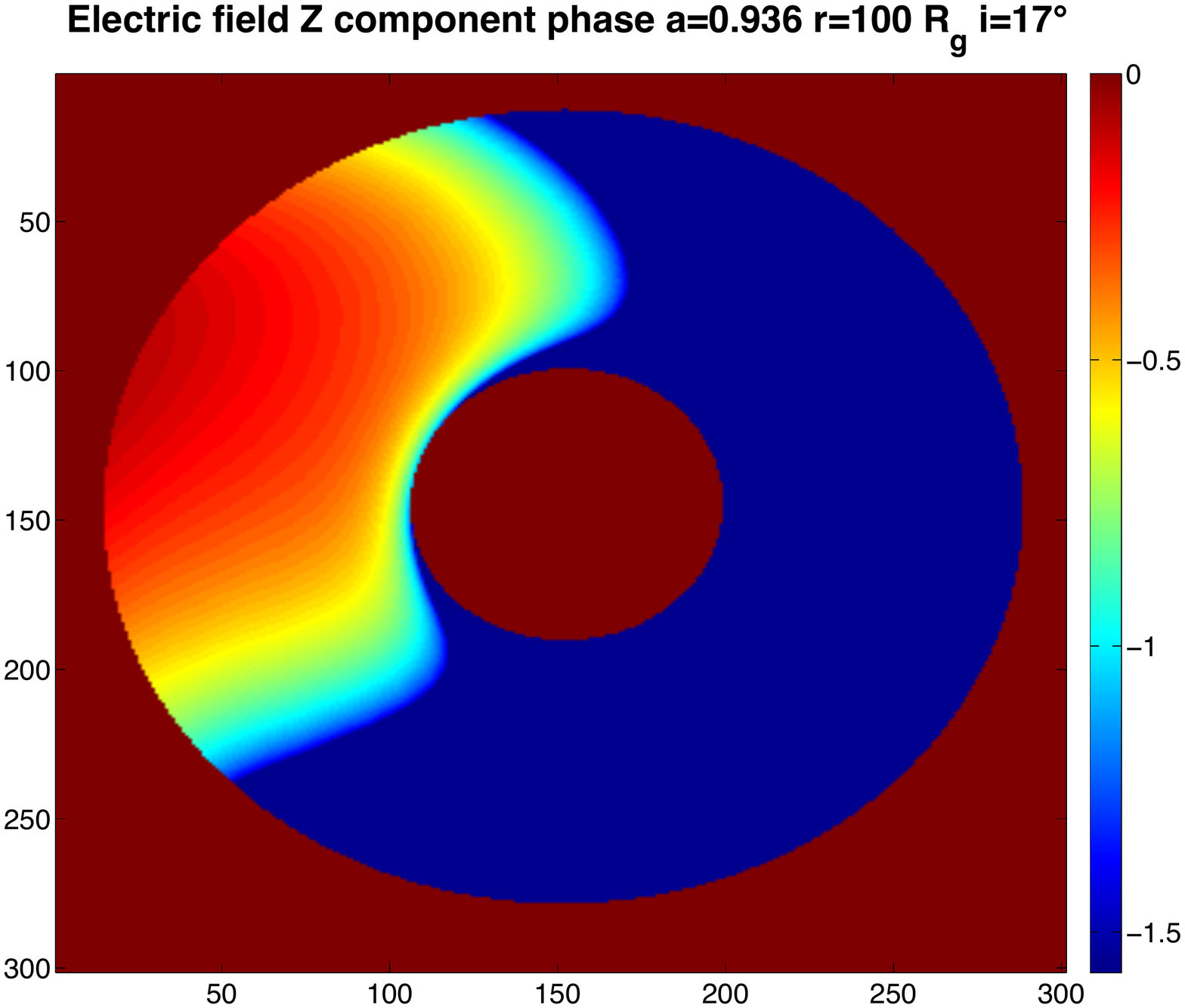}
\includegraphics*[width=.4\linewidth]{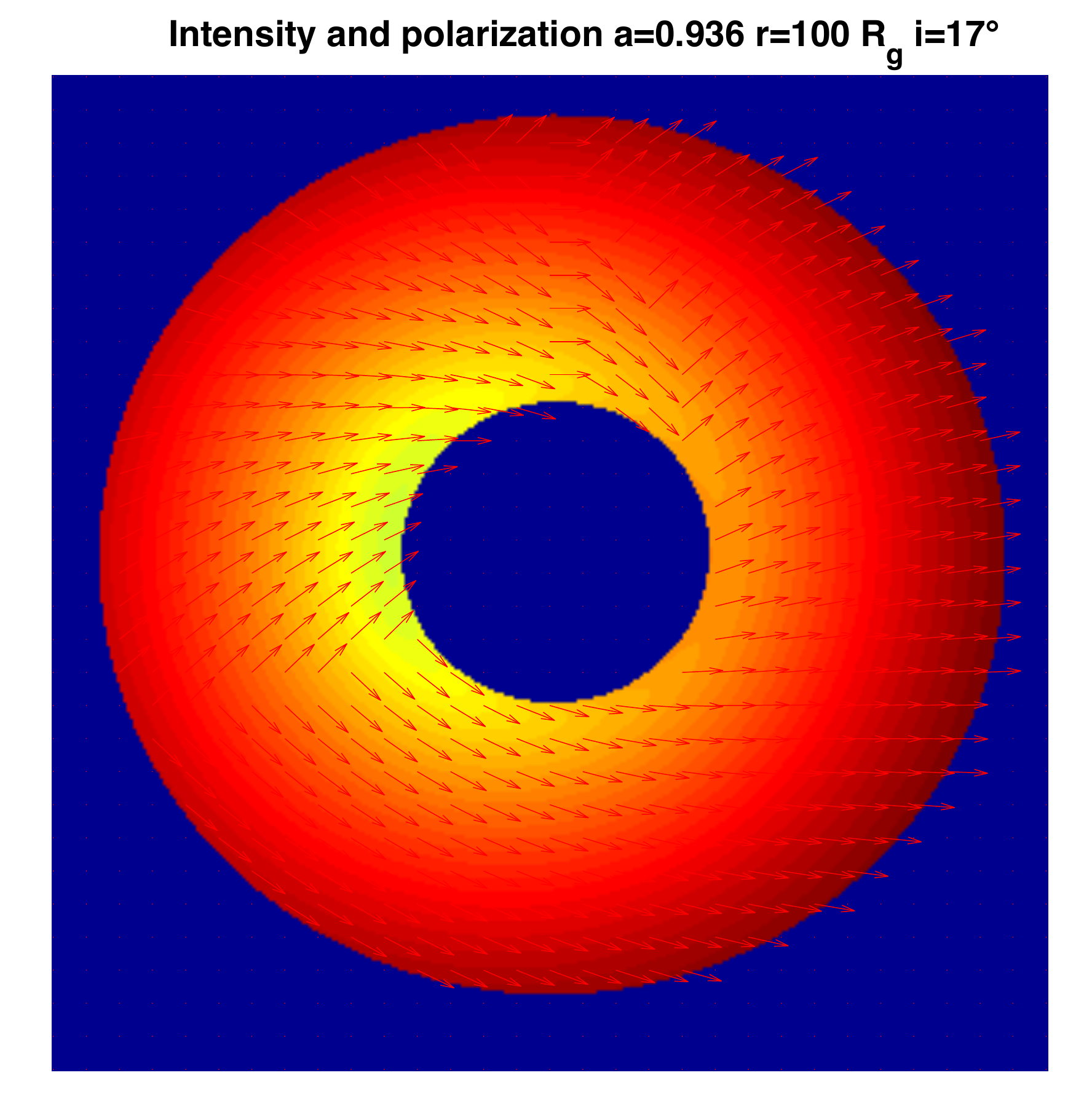}%
\includegraphics*[width=.5\linewidth]{sszt17.pdf}
 \caption{%
  \textbf{Results of numerical simulations from KERTAP} 
  ~\textbf{Upper panels:} Intensity, normalised to unity (left), 
  and phase (right) of the $z$ component of the radiation field emitted from
  the black hole accretion disc dominated by thermalised emission
  and synchrotron radiation.  The black hole has a spin pointing
  away from Earth and an inclination between the approaching jet and
  the line of sight of $i=17^\circ$. This is equivalent to a similar
  geometry with an inclination $i=163^\circ$ for a magnetic arrested
  disk \citep{vanp1,vanp2}, but where the spin is aligned with the
  accretion flow pointing toward the Earth. Polarization and spiral
  spectrum plots are shown in the lower panels.  This geometry, shown here
  at an orientation of $90^{\circ}$, agrees with that described by the
  experimental spiral spectra showing clockwise rotation as discussed
  in the main text, in agreement with the asymmetry parameter $q$,
  the ratio between the $\ell=0$ and $\ell=1$ peaks (OAM amplitudes)
  of the experimental data. This is smaller than that generated
  by the theoretical model, showing lower coherence in the emission
  mechanisms. The error is on the order of $10^{-7}$ and errorbars are
  not reported here.
 } \label{figkertap}
\end{figure*}

The OAM of the EM field emitted from the M87* surroundings is
induced by the gravitational lensing of the rotating black hole and
its rotation.  To characterize the rotation parameter obtained from the
analysis of the EHT data, we numerically simulate the gravitational lensing
around different BHs with different inclinations and rotation parameters.
Then we reconstruct the spatial phase profile from the Stokes parameters
(polarisation) provided by KERTAP and from the calculation of the
Pancharatnam-Berry phase.  An example is shown in figure~\ref{figkertap}.

In the paraxial approximation, the electric field vector
of the received beam can be written as [\cf Eq.~(1.5) in
Ref.~\citep{Allen&Padgett:Incollection:2011,Zhang&al:SR:2015}]
\begin{align}
\label{eq:E(x,y)}
 \E(x,y) =
  \mathrm{i}
  \omega\left[
   u_x\xunit + u_y\yunit + \frac{\mathrm{i}}{k}
   \left(\pdd{u_x}{x}+\pdd{u_y}{y}\right)\zunit
  \right]\exp(\mathrm{i}{}kz)
\end{align}
where $u_x$, $u_y$ and $u^*_x$, $u^*_y$ represent the complex amplitudes
of the $x$ and $y$~components and their complex conjugates, respectively,
of the electric field in the image plane. The quantity $\omega$ is the angular
frequency, and $k$ is the wave number.

In SI units the $z$ component of the EM orbital angular momentum density
is \citep{Thide&al:ARX:2014}
\begin{align*}
 \epz[\x\cross(\E\cross\B)]\bdot\zunit
 = \frac{\epz}{2\omega}\sum_{i=1}^{3}\cc{E}\hat{L}_zE_i
 = \frac{\epz}{2\omega}\sum_{i=1}^{3}\cc{E}_i\left(-\mathrm{i}\pdd{}{\ph}\right)E_i
\end{align*}
where $\x = x\xunit + y\yunit + z\zunit$ is the radius vector from the
source to the observer. Then, one obtains the $z$ component of the EM
angular momentum, expressed in terms of the complex amplitudes of the $x$
and $y$ components of the field as \citep{Zhang&al:SR:2015}
\begin{align}
\label{eq:Lz_parax}
 \hat{L}_z =
  \mathrm{i}\frac{\omega}{2}\left(
   u_x\pdd{u^*_x}{\ph}+u_y\pdd{u^*_y}{\ph}-u^*_x\pdd{}{\ph}-u^*_y\pdd{}{\ph}
  \right).
\end{align}

Taking the Stokes parameters $(I,U,V)$ for the linear and
elliptic polarisation polarisation states as reference fields, the
Pancharatnam-Berry phase of the vortex field $\E(x,y)$ that we want to
determine is given by the quantity \citep{Zhang&al:SR:2015}
\begin{align}
 \psi_{P(V/U)} = \arg\big(\langle\Phi_V\vert\Phi_\E\rangle\big)
\end{align}
More precisely, $\psi_{P(V/U)}$ is calculated with the argument of
the ratio between the circular/elliptic state of polarisation of the
EM field, characterized by the Stokes parameter $V$, and the state
${\left|\Phi_V\right\rangle}$, and that of the initial field $\E(x,y)$,
namely, ${\left|\Phi_\E\right\rangle}$.

According to \citet{Zhang&al:SR:2015}, one obtains the average value of
the OAM topological charge $m$ of the field, for any concentric circle of
pixels having radius $r_c$ and with the origin at the center of the image,
from
\begin{align}
\label{eq:dl/dr}
 \pdd{m}{\ph} =
 r_c I_N\left(
   -I\pdd{\psi_{P(V/U)}}{\ph}\mp(I\pm U)\pdd{\psi_s}{\ph}
  \right).
\end{align}
This quantity is implemented numerically in the plane of
observation of an asymptotic observer. In Eq.~\ref{eq:dl/dr}
the quantity $I_N$ is  the normalised intensity.  The first term,
$I\,{\partial{(\psi_{P(V/U)})}}/{\partial\ph}$, is the gradient of the
spiral spatial phase distribution obtained from the Stokes parameters $V$
and $U$.  Finally, the term $(I\pm U)\,{\partial\psi_s}/{\partial\ph}$,
describes the variation of the state of the polarisation in space
at each point in the observational plane of the asymptotic observer.
At the end one can calculate the spiral spectrum.

\subsection{Reconstructing the spatial phase distribution from different
intensity acquisitions}

Here we describe how we determined the rotation of the M87*
black hole from the brightness temperature distributions
\citep{Rybicki&Lightman:Book:RadiativeProcesses:2004} provided
by EHT, by using the Transport of Intensity Equation (TIE) method
\citep{Barbero:OENGR:2006,Schulze&al:OE:2012,Lubk&al:PRL:2013,Ruelas&al:JO:2018,Kelly:IJO:2018},
adopting a finite-difference method for a series of matrices
($250\times250$) of intensity values.

As described in the standard literature, the OAM of the EM field
can be determined from the spatial phase distribution and the spatial
intensity (or amplitude) distribution in the plane of observation
of an interferometer \citep{Tamburini&al:APL:2011}.  If the spatial
phase distribution is not available it can be reconstructed with
the TIE method.  To explain this procedure better, let us consider
the intensity of the EM field as in Eq.~\ref{eq:E(x,y)}; the
EM waves are described in terms of complex-valued terms of the
amplitude $u(r)$ in the following paraxial wave equation
\citep[see]{Goodman:Book:IntroductionFourier:2016,Ruelas&al:JO:2018}
\begin{align}
\label{eq:parax_wave}
 \left(\del_{(x,y)}\bdot\grad_{(x,y)}
  +4\pi\mathrm{i}\pdd{}{z} \right)u(r) = 0
\end{align}
where $\del_{(x,y)}$ represents the vector nabla differential operator
in the image plane $(x,y)$ perpendicular to the $z$ axis connecting the
BH and the asymptotic observer.

By multiplying equation~\eqref{eq:parax_wave} with its complex conjugate,
one can split this equation into a system of partial differential
equations (the system of TIE equations) that describe the intensity and the
phase evolution with respect to the translational motion across the $z$ axis as
explained in~Ref.~\citep{Ruelas&al:JO:2018},
\begin{subequations}
\begin{align}
\label{eq:dI/dz_dP/dz}
 \pdd{I}{z}
&= -\frac{1}{2\pi}\del_{(x,y)}\bdot\left(I~\del_{(x,y)}P\right)
\\
 \pdd{P}{z}
&= \frac{1}{4\pi\sqrt{I}}\nabla^2_{(x,y)}\sqrt{I}-\lvert{\del_{(x,y)}P}\rvert^2
\end{align} 
\end{subequations}
Here, $I$ is the intensity and $P$ the phase distributions in the plane
of propagation that evolves along the $z$~coordinate. 
We solve numerically this system of equation point by point in the
image plane, for given values of $I$, and recover $P$.  The evolution of
the phase $P$ and intensity $I$ along the $z$~axis -- connecting M87*
and the Earth -- is provided by the relative motion of the Earth with
respect to M87 occurred between each different acquisition.

The quantity $\lvert{\del_{(x,y)}P}\rvert$ is the magnitude of the
phase gradient vector in the image plane.  To summarise, by using the
numerical solution of the system of equations \ref{eq:dI/dz_dP/dz} in the
image plane of an asymptotic observer, we reconstruct the spatial phase
distribution from the intensity plots provided by EHT and determine the
rotation from the analysis of the OAM spiral spectrum.  The asymmetry
parameter $q$ given by the ratio between the $\ell=1$ and $\ell=-1$
components (amplitudes) of the OAM spiral spectrum the histogram is
compared with the values of $q$ obtained from the numerical simulations
performed with KERTAP.  Then, using Matlab, we obtained the rotation
parameter $a$ by a polynomial interpolation of $a$ as a function of
$q$ obtained by varying the inclination parameter $i$ for any type of
accretion disk.  As discussed also by the EHT team in their numerical
simulations \cite{EHT6}, we also find that the image properties are
determined mainly by the spacetime geometry, that are dominant, and then
by the AD characteristics.

\subsection{Applicability of the TIE method}

As discussed in the main text, the EM waves at $230$ GHz propagate
unaffected by the scattering, e.g. due to the interstellar medium
and other astrophysical effects \citep{Tamburini&al:AA:2011} from
M87* to the Earth.  The propagation of these EM waves remains quite
stable also during the relative motion of the source with respect to
the observer during the time interval of one day that separates two
consecutive intensity acquisitions in each of the two acquisition epochs.
To give an estimate of the relative shift between M87* and the receiving
radiotelescopes, one has to consider that the relative motion of M87 is
due to an heliocentric radial velocity $v_\text{rad}=1282\pm7$~km/sec 
and the source is located at a distance of $54.8\pm2.6\times10^6$ light years
(ly) \citep{10.1111/j.1365-2966.2010.18174.x,ned}.  The ratio between the total propagation distance from M87* to
the Earth and that due to their relative motion in one day is very small
\begin{align}
 \frac{\Delta{L}_{\text{day}}}{\Delta{L}_{\text{tot}}} \approx
 2.5\times10^{-12}.
\end{align}
The change of the EM wavefront from M87* in one day of motion is then
due mainly to the phase profile of the wavefront, namely due to the
presence of OAM.  In fact, the propagation of EM waves at that frequency
cannot be affected by possible astrophysical effects in the spatial
interval traveled by the Earth with respect to M87* during one day.
The explanation is quite trivial: this distance is extremely
small with respect to the total distance from M87* and our planet and
there are not extreme astrophysical phenomena around the Earth's orbit
that can affect the propagation of light \citep{Tamburini&al:AA:2011}.
The small deviations observed are due to acquisition/instrumental errors
and from the unknown remainder modulo $\lambda$ separating the two
acquisitions of each of the two observational epochs.  This quantity
can be handled as a stochastic term in the determination of $a$.
The mathematical validity of our procedure based on the TIE method is
ensured by the Wold theorem described below.

\section{The Wold Theorem}

Wold's decomposition (or representation) theorem states that every
covariance-stationary time series $Y_{t}$ or, equivalently, the time
evolution of a discrete process in time, can be decomposed into a sum
of two time series: one deterministic time series and a stochastic one.

The time-discrete process that we consider here is represented by the
evolution of the brightness temperature value $Y_t$ in each image
pixel at the position $(x,y)$ of the BH image and at the time $t$
of observation.  The evolution of $Y_t$ is a time series that can be
described in terms of covariance-stationary time series because the
physical variations observed in M87* are very small, as reported in
Refs.~\citep{EHT5,Doeleman&al:SCI:2012}.  By applying the Wold theorem
this process can effectively be approximated and modeled as a sum of
a deterministic process plus a stochastic deviation, in which all the
terms of the sequence have the same mean, and the covariance between any
two terms of the sequence depends only on the relative positions of the
two terms.

The deterministic part represents the slow temporal evolution of M87*,
which is a quasi-static source \citep{EHT4,EHT5,EHT6}, when observed in
the intervals of one day that separate the acquisitions of each of the two
pairs of nights (5--6 April and 10--11 April) with a broad consistency.
The stochastic deviation from the deterministic part, on the other hand,
is given by the contribution of the undetermined variation modulo the
wavelength $\lambda$ that occurs during the long translational motion
of a terrestrian observer with respect to the source, M87 and from
the experimental errors.

The phase is reconstructed with the help of the TIE method, which is
always a deterministic process that approximates an ideal time series
of events described in the interval $0\leq{t}<\infty$ by the sequence
\begin{equation}
\label{eq:Wold}
 Y_t =
  \sum_{j=0}^{\infty} b_j \epsilon_{t-j} + \eta_t
\end{equation}
where $\epsilon_t$ is an uncorrelated sequence.  The innovation process
$\epsilon_t$ \footnote{The innovation process is the difference between
the observed value of a variable at time $t$ and the optimal forecast
of it, based on information available prior to $t$. If the representation
of the stochastic process is correct, any successive innovation results are
uncorrelated and innovations are modeled with a white noise time series.}
can be represented by a white noise process. The terms $b_j$ act as 
linear filters and are constant in time. The vector $b$ is essentially
the coefficient vector of the stable stochastic process, and $\eta_t$
is the information encoded in the images, namely, the field amplitude
or the intensity in each of the acquisitions.  Any stationary process
such as our phase reconstruction process can be described with a good
approximation by this particular representation.

The dynamical evolution of any process with these properties can be
approximated by a linear model and if all the stochastic processes are
independent, this is the only linear representation possible of this
process, especially when the stochastic process is small with respect to
the deterministic process.  As described by \citet{Ruelas&al:JO:2018},
the TIE analysis can unambiguously extract all the information encoded in
the two (or more) different images.  In fact, introducing a stochastic
noise in the variable $u$, and thus in the intensity $I$ the system of
equations~\ref{eq:dI/dz_dP/dz} becomes equivalent to a system of
stochastic differential equations that can be solved numerically in a
domain where the sum of the effective value of the brightness temperature
and of the stochastic process are different from zero. This condition
is always satisfied in the values of the brightness temperature in the
EHT images of M87*. This proves the mathematical consistency of the TIE
method in our approach of estimating the spatial phase distribution.

\section{Spiral spectra and BH rotation: Additional material}

To proceed with the analysis of the BH rotation one must obtain a
reference OAM (spiral) spectrum from a set of numerical simulations
of an Einstein ring.
Then, one has to characterize the OAM content in the simulated image
from the analysis of the amplitude (or intensity) and the phase of the
field in the image plane of an asymptotic observer.  As explained in
the main text, the asymmetry of the OAM states $\ell=1$ and $\ell=-1$
indicated by the heights in the columns of the histogram of the spiral
spectrum allows us to estimate, with good accuracy, the twist in the
light due to the rotation of the BH.

The ratio $q$ between the height of the $\ell=1$ and $\ell=-1$ column
peaks (OAM amplitudes) in the spiral spectrum for an Einstein ring as
observed in M87* if, in the simulations, we assume a photon power law
index in the range $0.1<\Gamma<2$. We concentrate our efforts on the
canonical case \citep{Falcke&Biermann:AA:1995} $\Gamma=2$ and assume a
radial power law index that specifies the radial steepness of the profile,
namely, $n_r=3$.

\begin{figure*} \centering
 \includegraphics*[width=.5\linewidth]{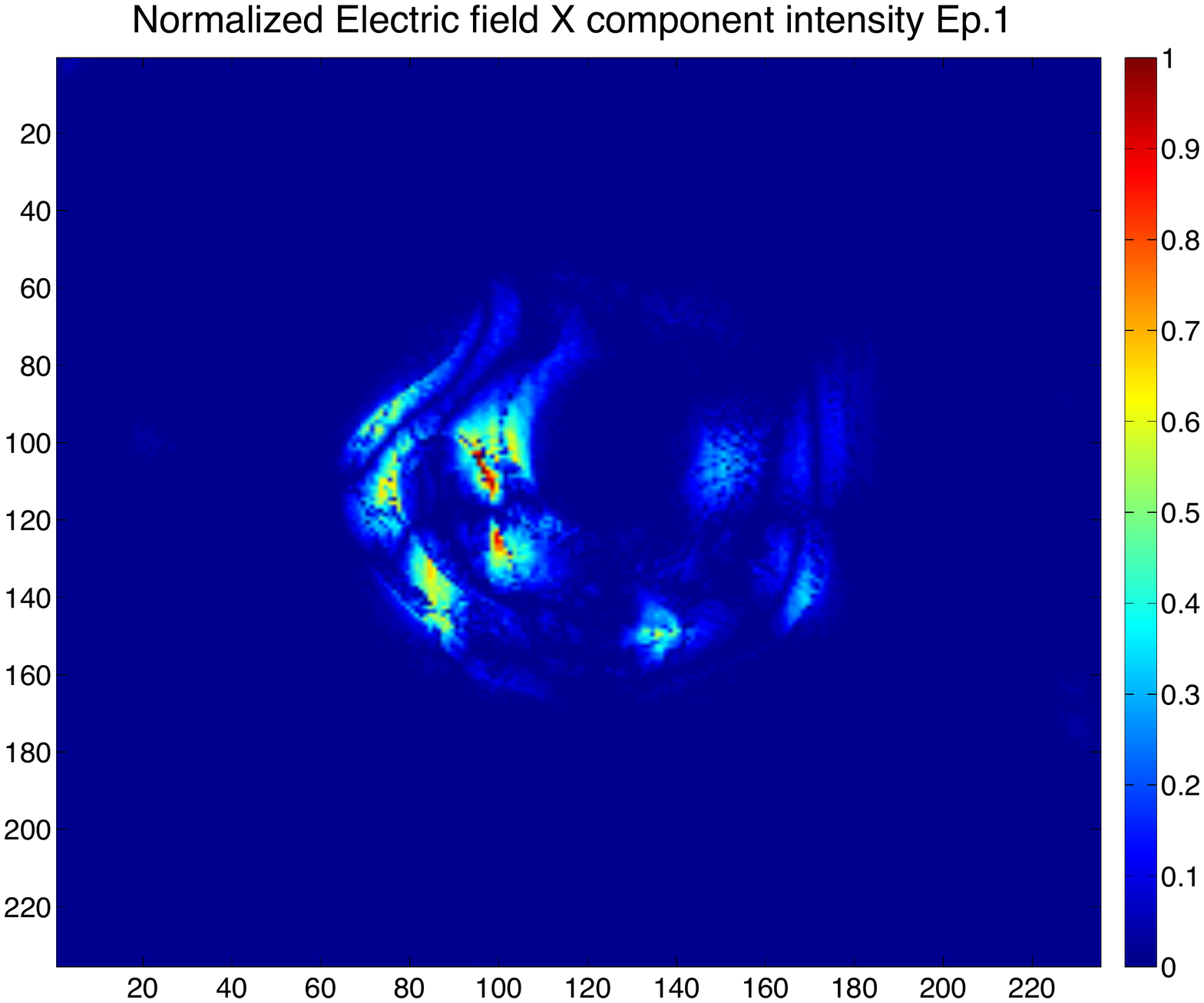}%
 \includegraphics*[width=.5\linewidth]{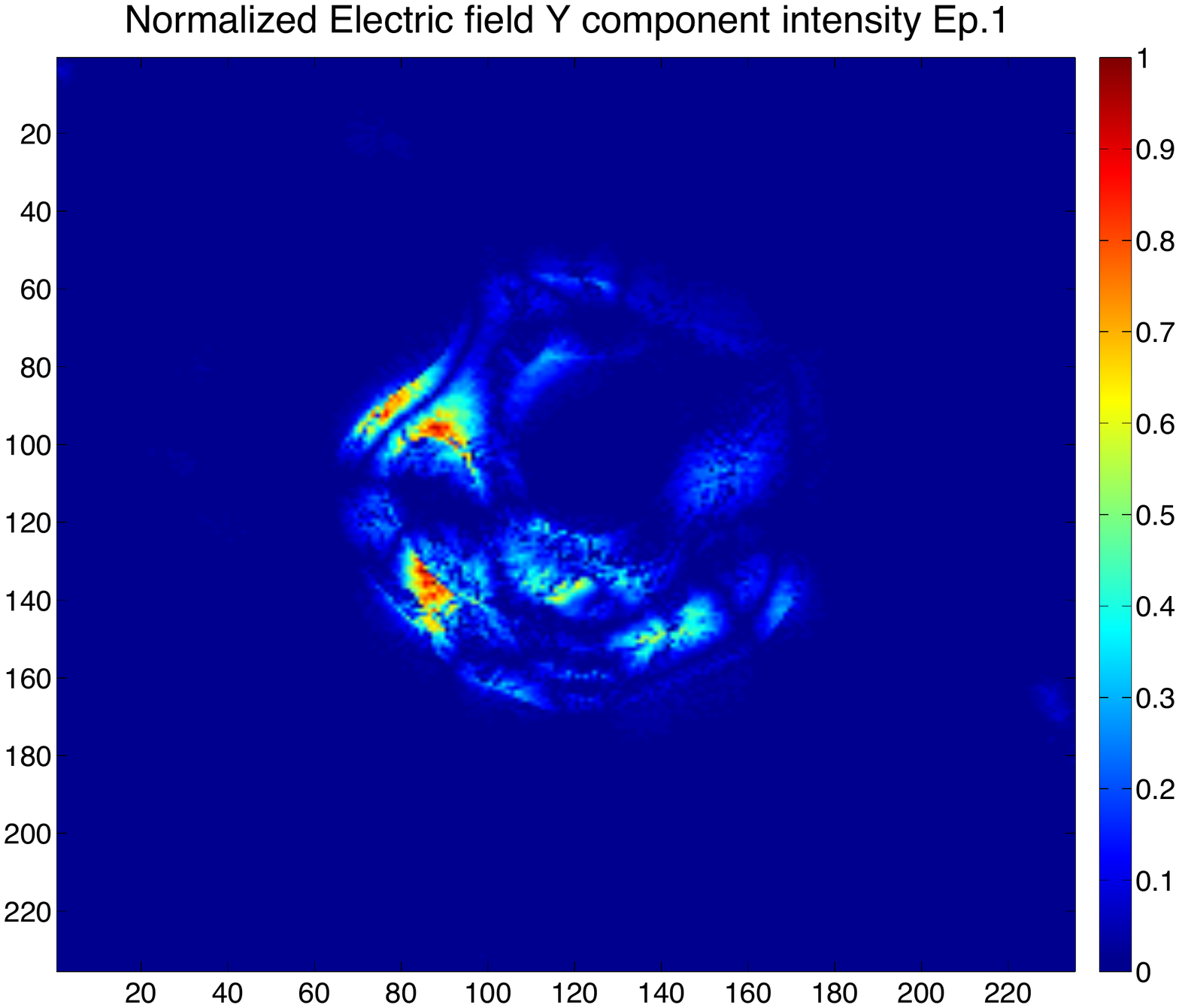}\\
 \includegraphics*[width=.5\linewidth]{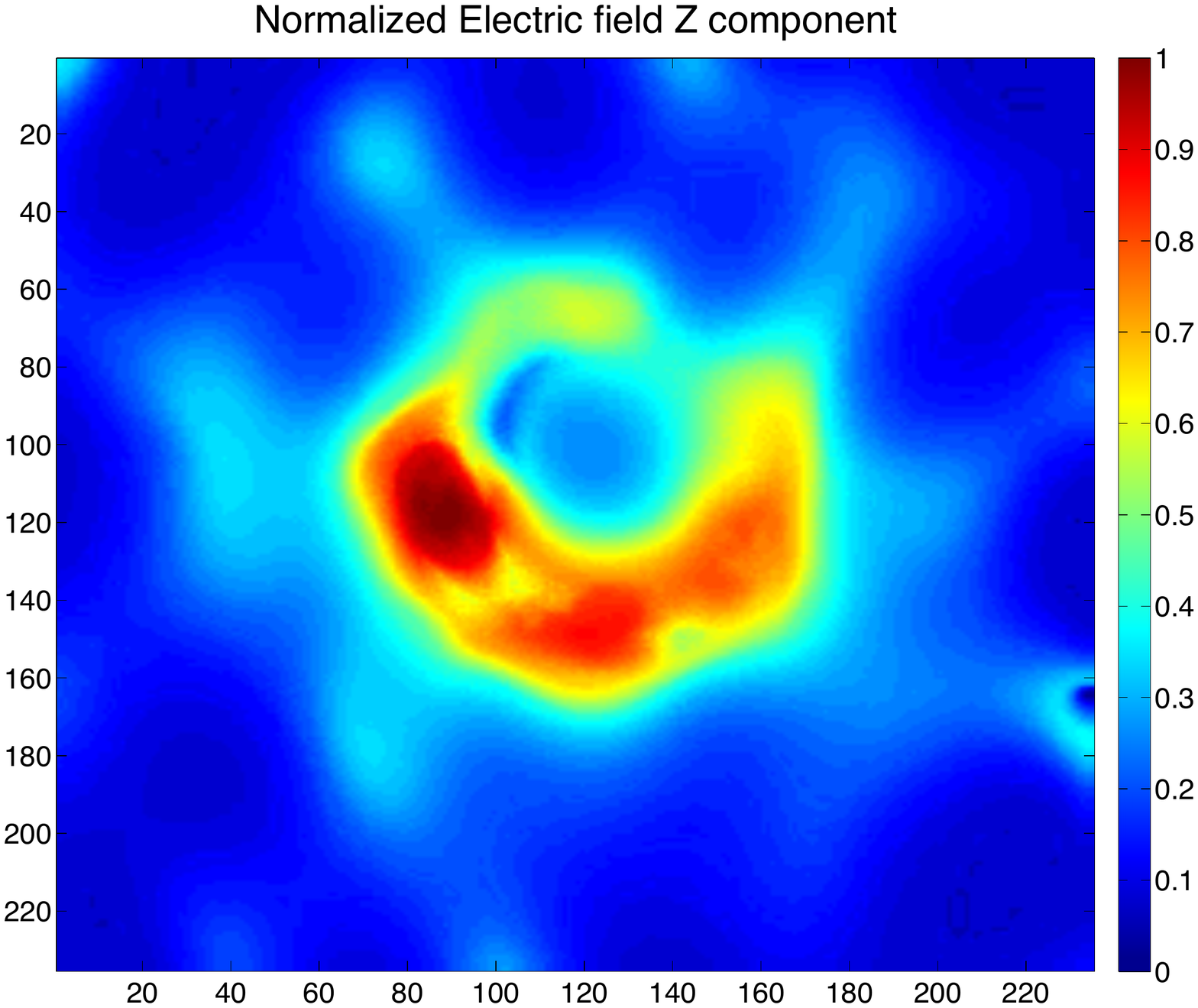}%
 \includegraphics*[width=.5\linewidth]{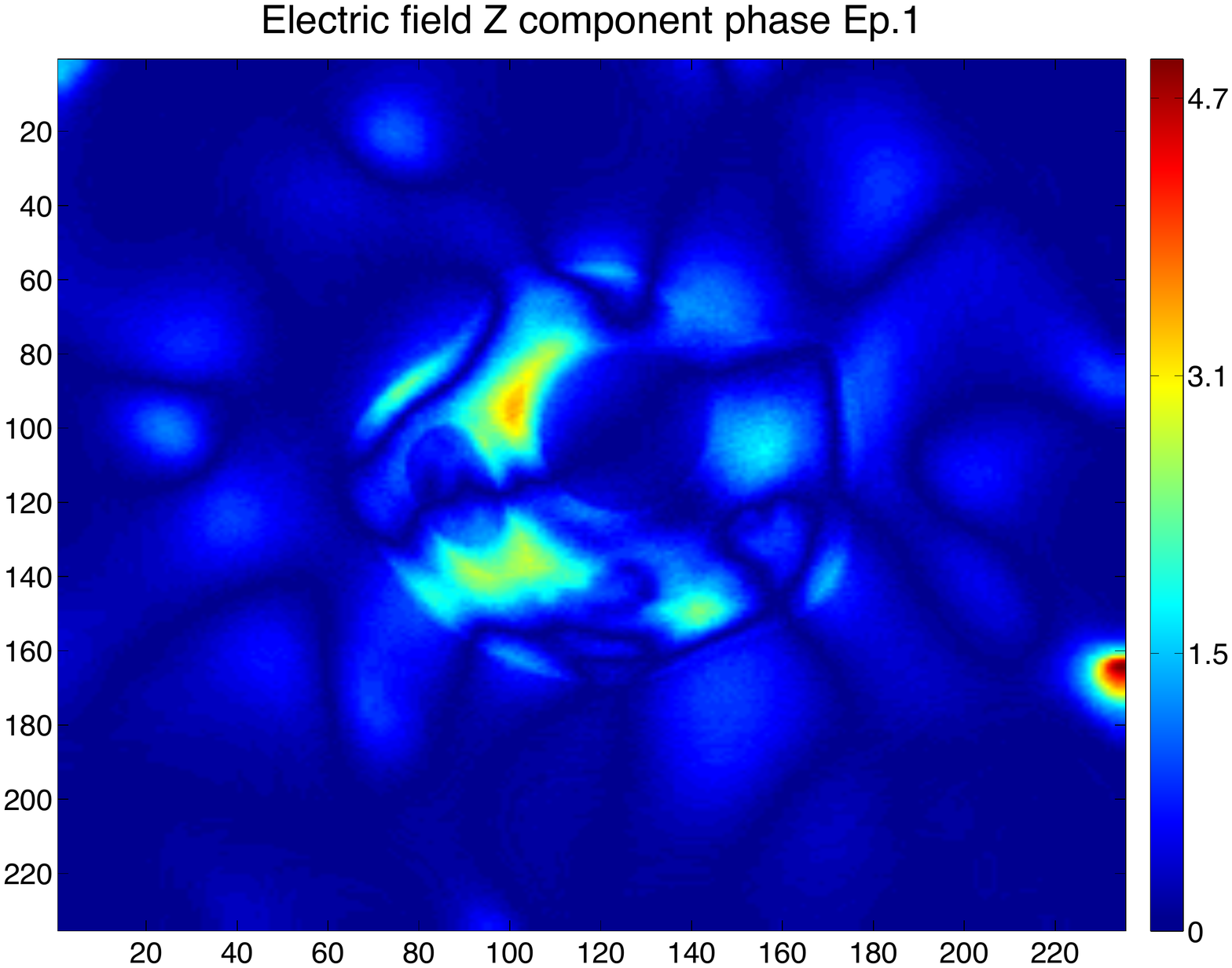}\\
 \includegraphics*[width=.5\linewidth]{SStieepoch1_errorbar.pdf}
 \caption{\textbf{Upper panels:} Normalised electric field $x$ (left) and
 $y$ component (right) reconstructed from the two EHT data acquisitions
 taken during epoch $1$ (April 5 and 6, 2017, respectively),
 with the Transport of Intensity Equation (TIE) method.  \textbf{Lower
 panels:} Normalised electric field intensity and phase across the
 direction of observation $z$ and the corresponding spiral spectrum.
 The spiral spectrum shows the presence of a black hole rotating clockwise
 with rotation parameter $a=0.90\pm0.10$ and inclination $17^{\circ}$
 pointing away from Earth (see main text).  The TIE reconstruction method
 of the fields is possible because the M87 black hole Einstein ring is
 spatially resolved by EHT. The wavefront reconstructed here is not plane
 and therefore different from that emitted from a point-like source at
 infinity. The EM fields can be described as an unguided transverse
 magnetic (TM) beam in free space \citep{Davis&Patsakos:OL:1981}.
 The image coordinates are in arbitrary units.  The intensity sidebars
 are normalised to unity. The phase sidebars are in arbitrary units.
} \label{fig:supp1} \end{figure*}

\begin{figure*}
\centering
 \includegraphics*[width=.5\linewidth]{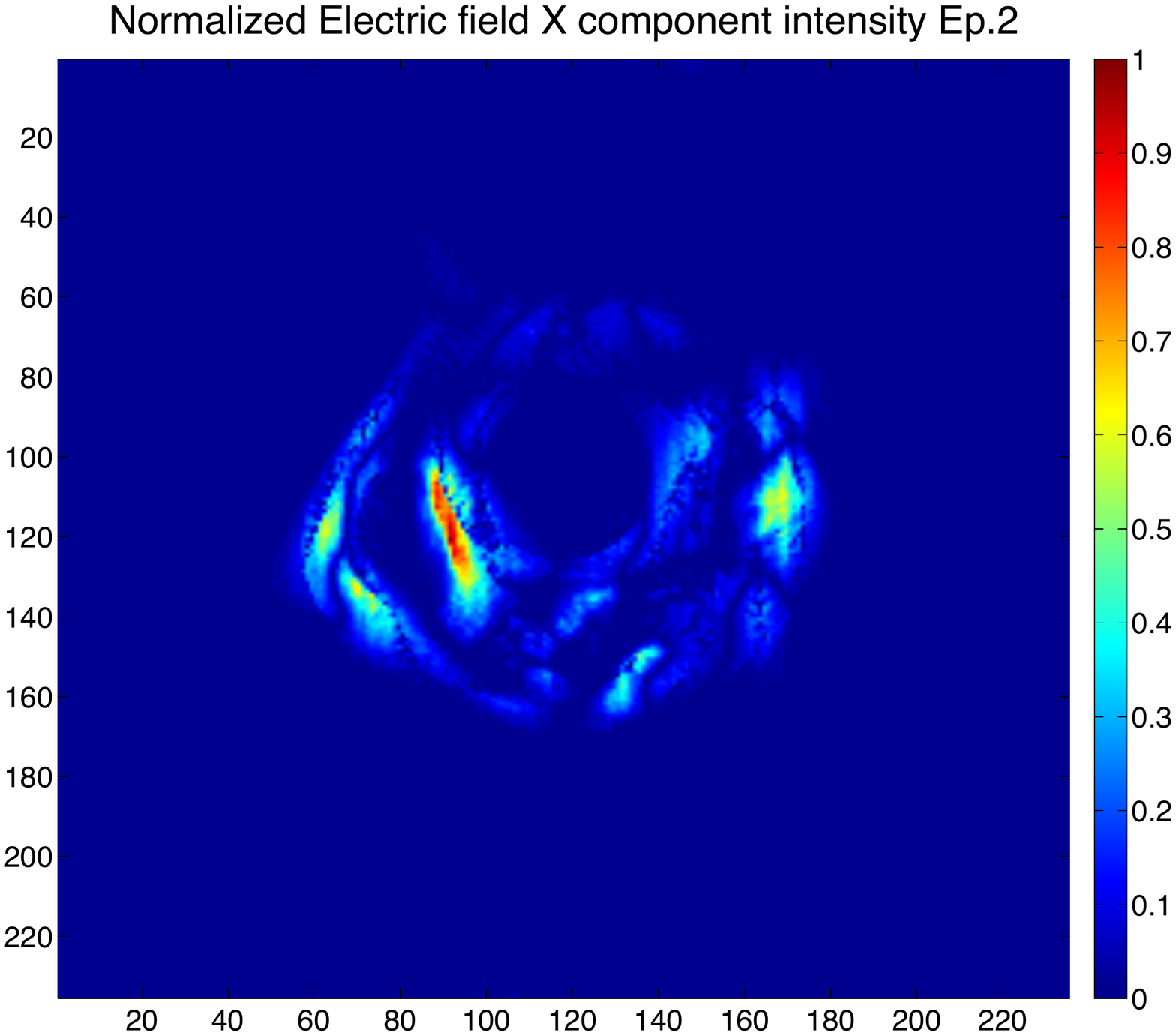}%
 \includegraphics*[width=.5\linewidth]{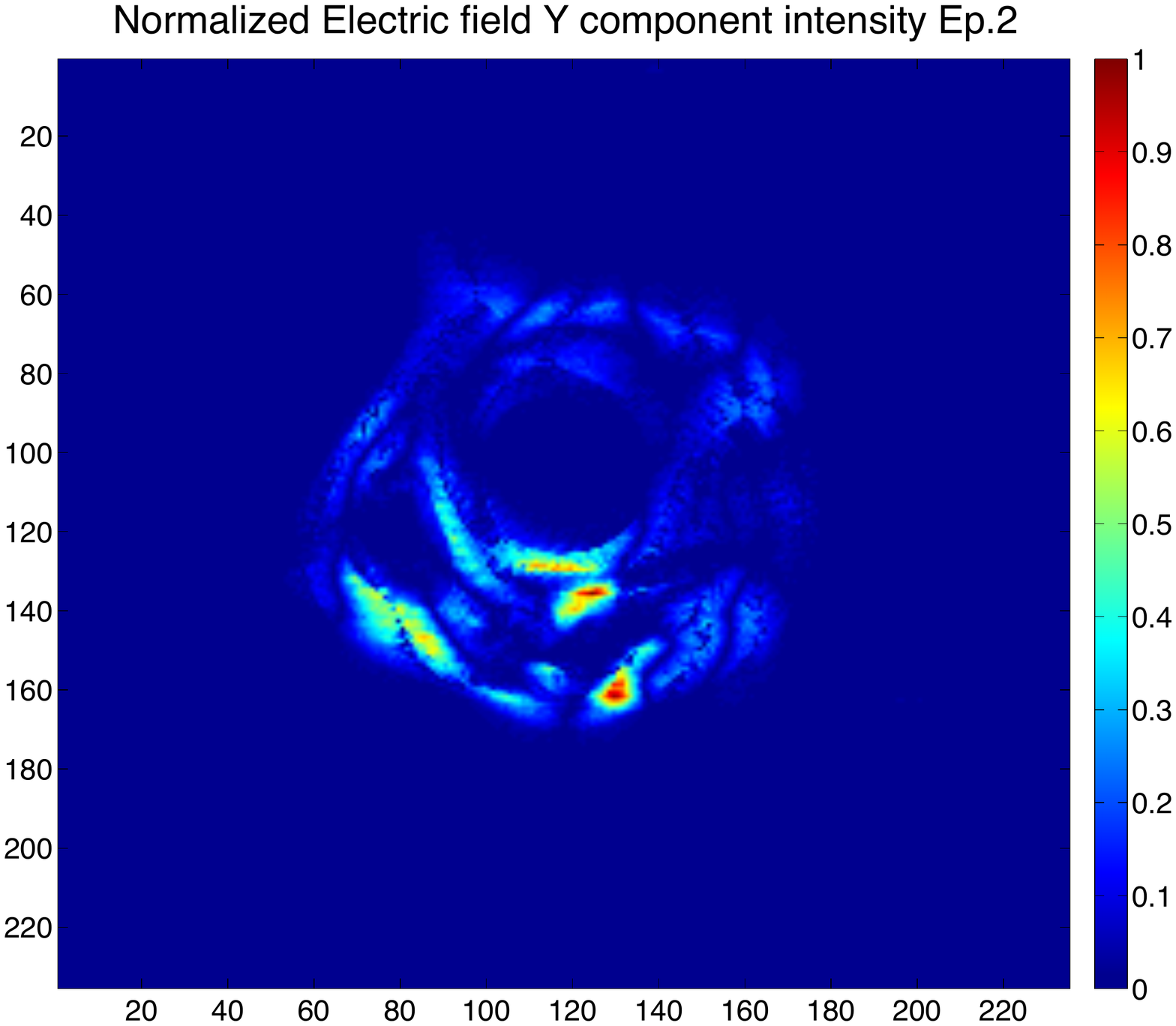}\\
 \includegraphics*[width=.5\linewidth]{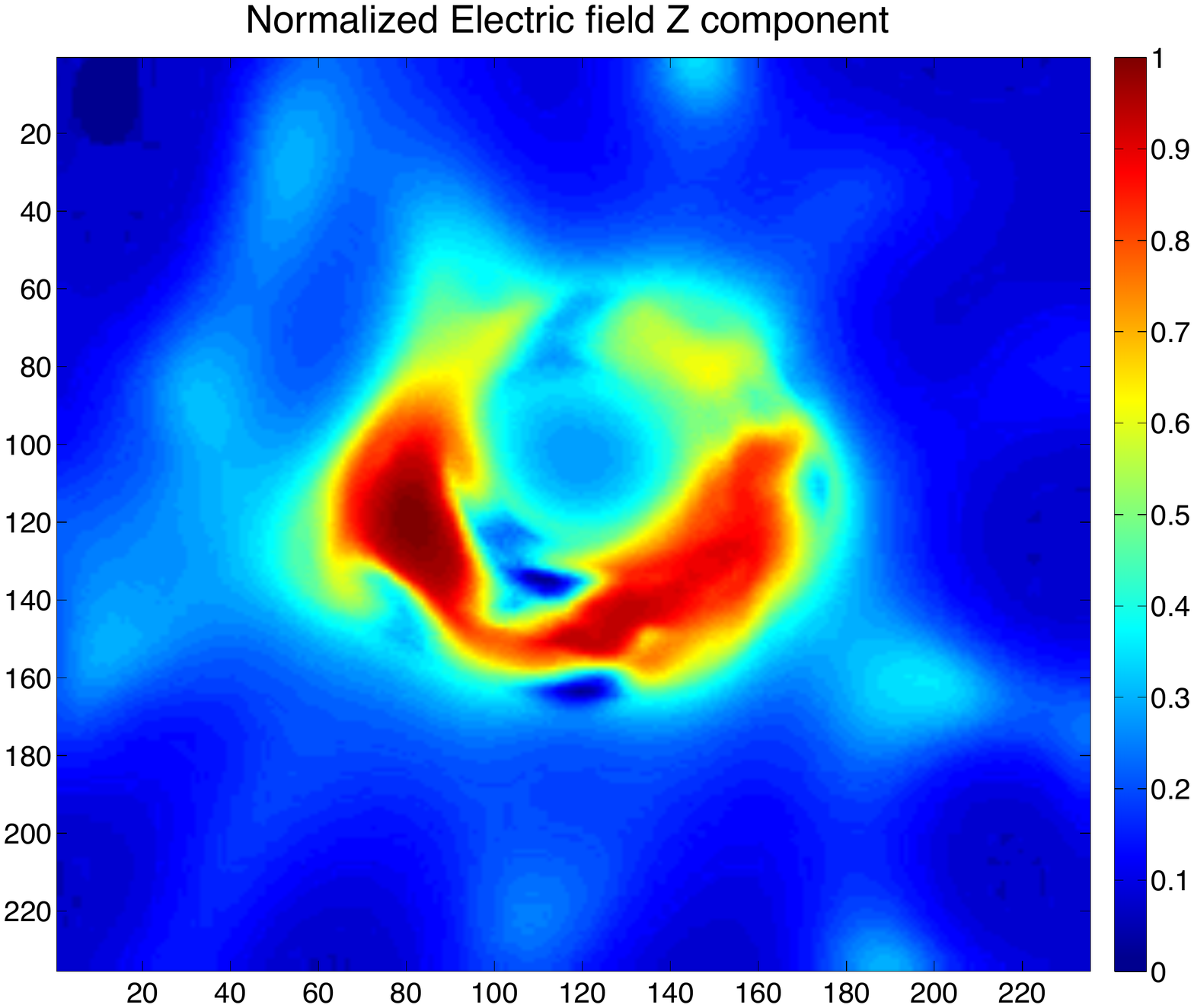}%
 \includegraphics*[width=.5\linewidth]{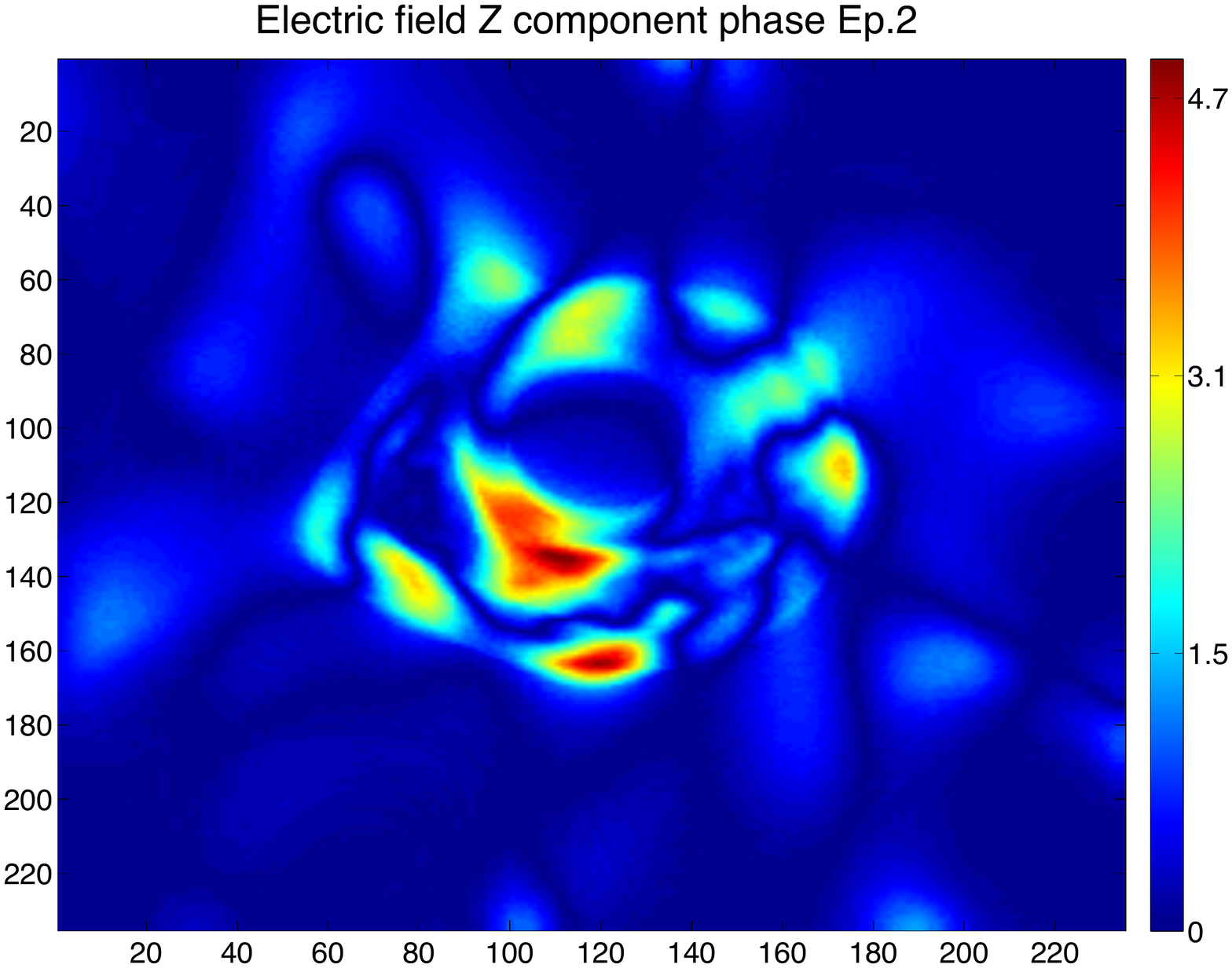}\\
 \includegraphics*[width=.5\linewidth]{SSTIEepoch2_errorbar.pdf}
 \caption{\textbf{Epoch 2.}
  Field components and spiral spectrum obtained with the TIE method
  as in SM Fig.~\ref{fig:supp1}. Also here, the spiral spectrum, as
  in epoch 1 that the component of the electromagnetic field along the
  propagation direction to the observer, agree with the results of a twist
  in the light caused by a clockwise rotating black hole with rotation
  parameter $a=0.90\pm0.10$ as previously discussed (see main text).
  The image coordinates are in arbitrary units.  The intensity sidebars
  are normalised to unity. The phase sidebars are in arbitrary units.
 }
 \label{fig:supp2}
\end{figure*}

We present the plots of the field amplitudes and the $z$ phase
for the observational epochs discussed in the main text, epoch~1 (SM
Fig.~\ref{fig:supp1}), and epoch~2 (SM Fig.~\ref{fig:supp2}).

\begin{figure*}
\centering
 \includegraphics*[width=.5\linewidth]{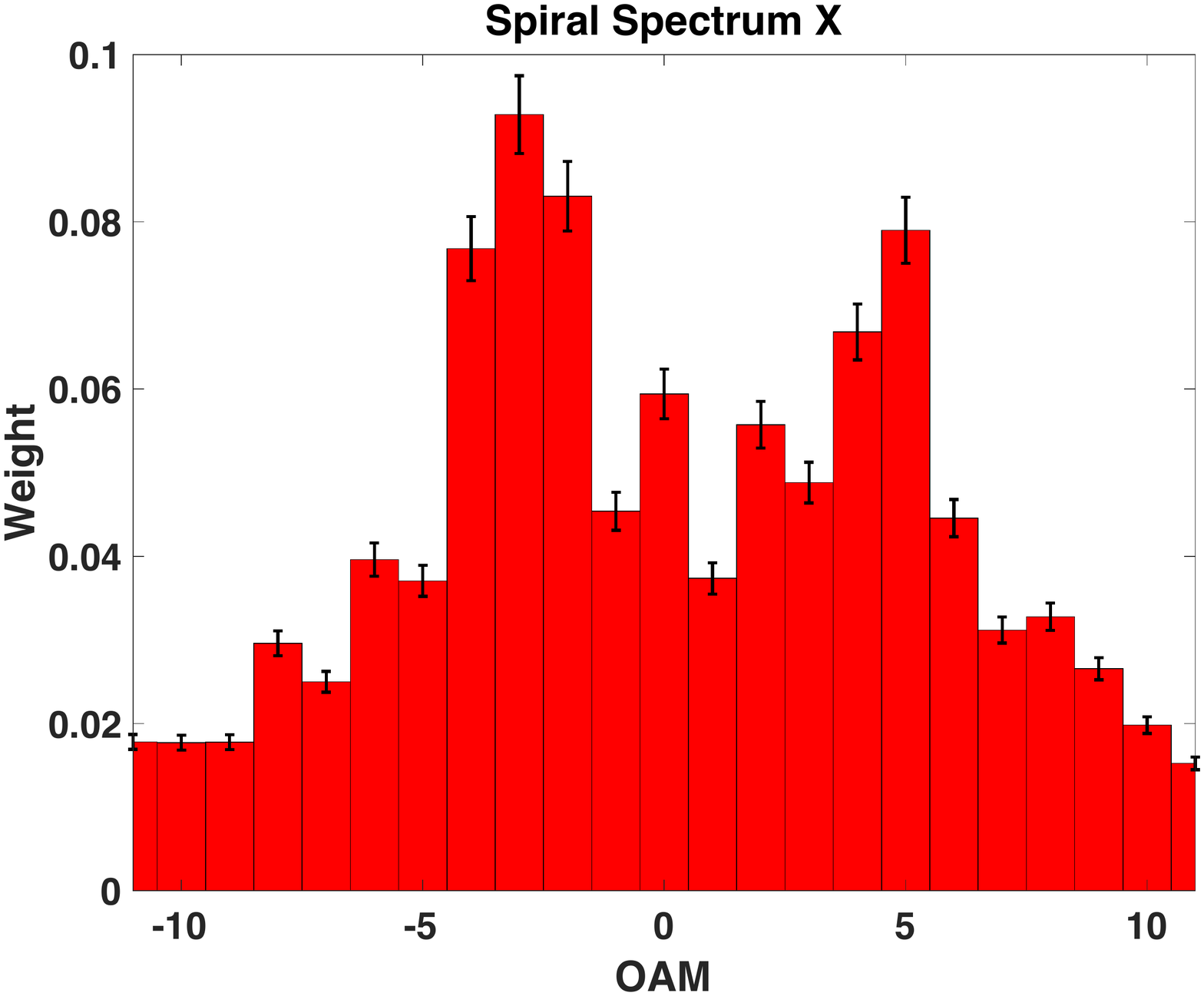}%
 \includegraphics*[width=.5\linewidth]{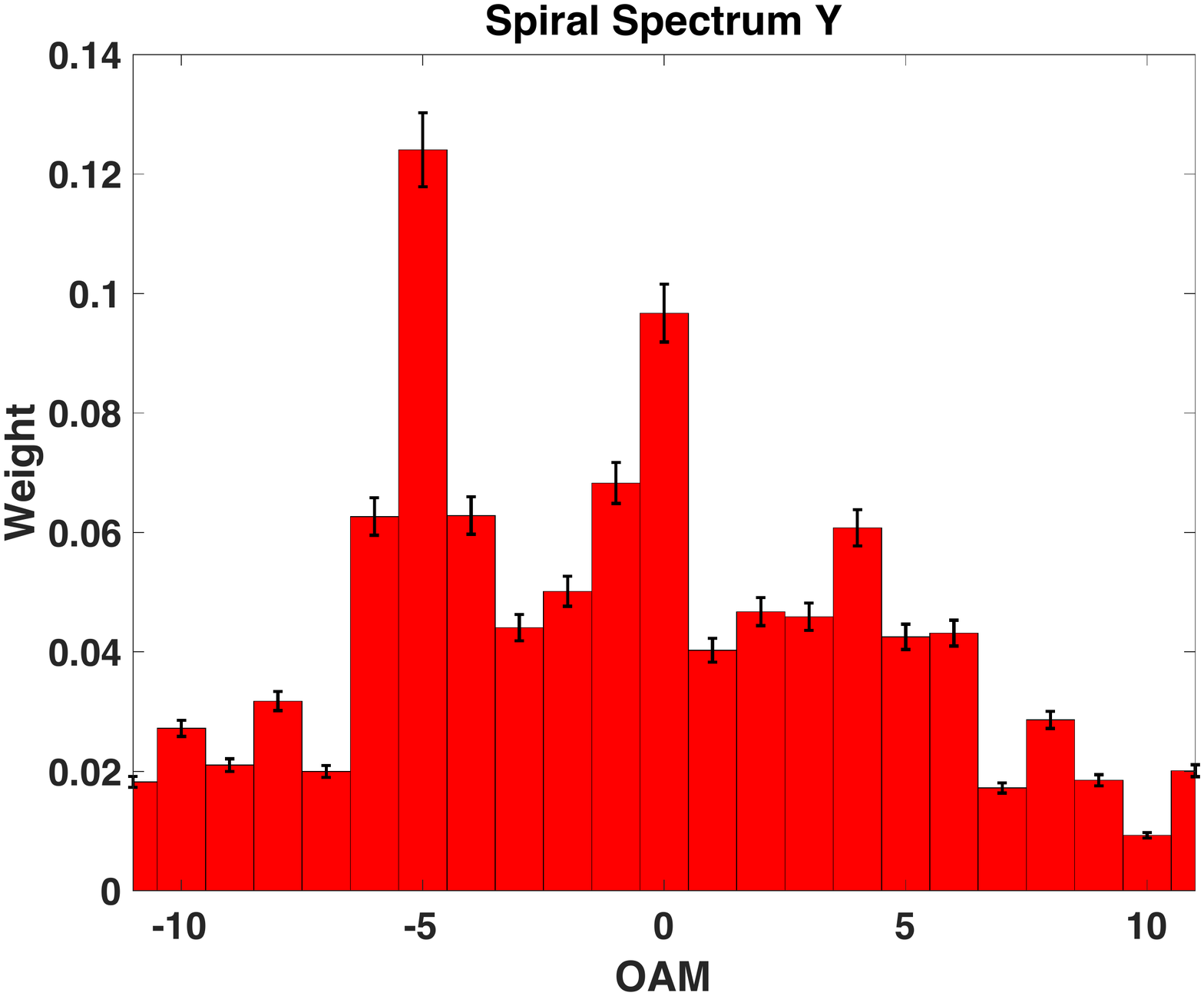}\\
  \includegraphics*[width=.5\linewidth]{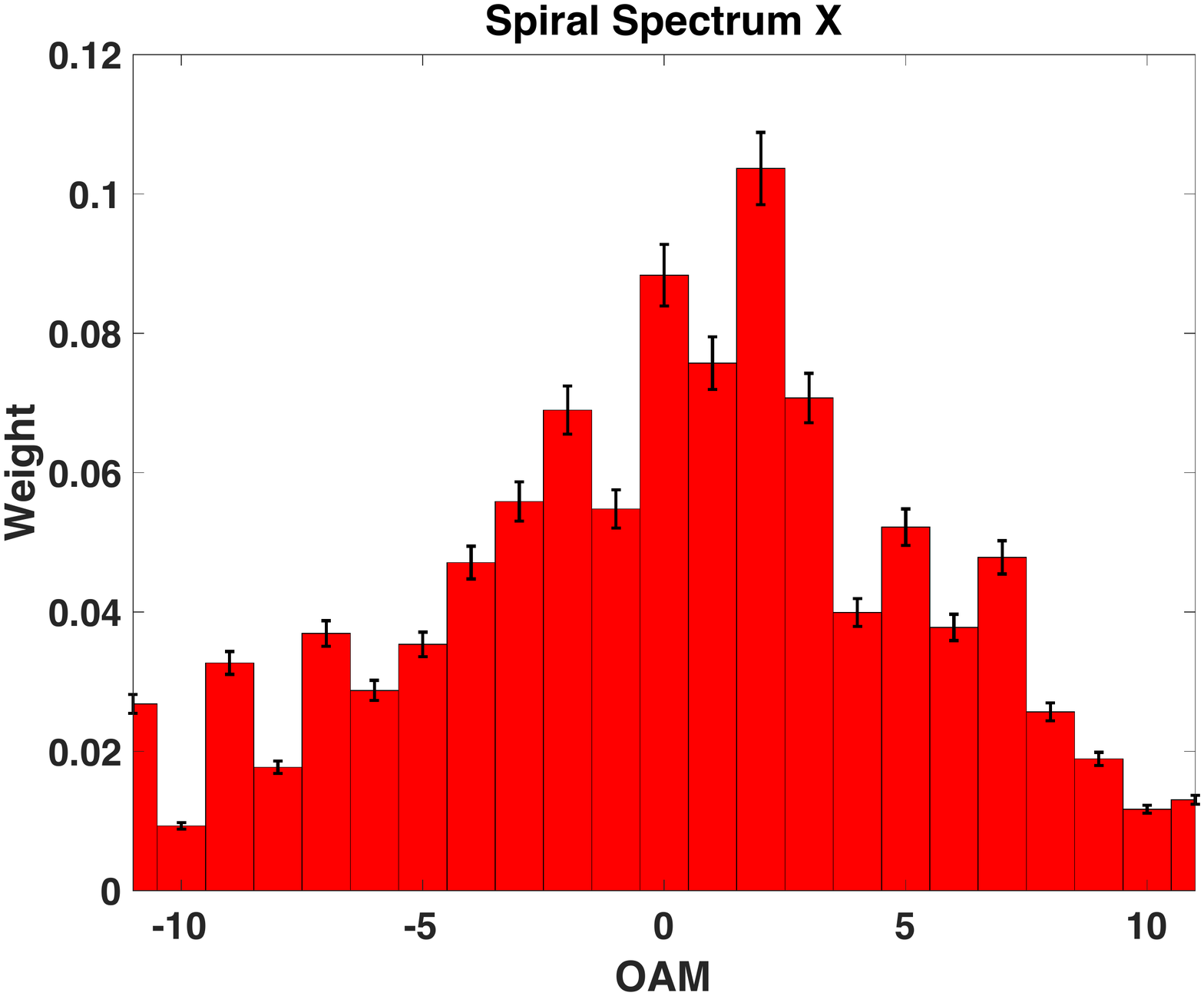}%
 \includegraphics*[width=.5\linewidth]{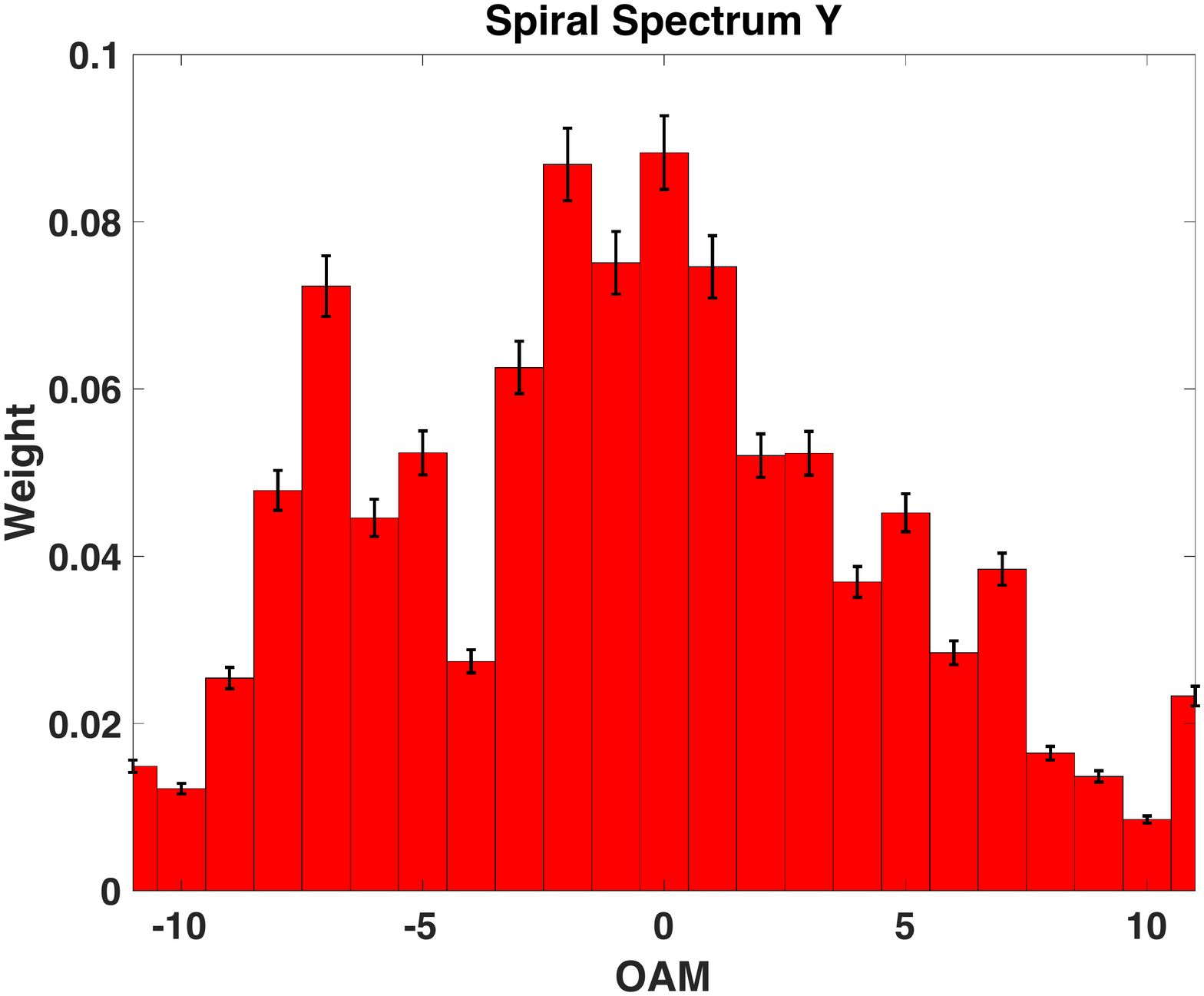}\\
 \caption{\textbf{Upper panel:}
  Spiral spectra of the $x$ and $y$ components of the electric field,
  obtained with the TIE method from EHT data in epoch 1 and those in
  epoch 2 (\textbf{Bottom panel}). Both components in both epochs do
  not show a predominant OAM component, as expected.
 }
\label{fig:supp3}
\end{figure*}

\subsection{Coherence of the EM waves and the TIE method}

In the spiral spectra plots of Fig.~1 and Fig.~2 in the main text, and
in Fig.~2, of the SM, we notice that the ratios between the $\ell=0$
and $\ell=1$ modes obtained with the numerical simulations are slightly
different from those obtained with the visibility intensity and phase
plots and different from those obtained from the TIE method.

If we describe the system with a theoretical model with a photon power law
index $\Gamma=2$ and a radial power law index $n_r=3$, one obtains the
amplitude ratio $q=h(\ell=0)/h(\ell=1)=7.6$, whereas the experimental
data analysed with the TIE method we find for epoch~1 (days April
$5$\textendash April $6$) that the ratio $h(\ell=0)/h(\ell=1)=14.65$
and for epoch~2 (days April $10$\textendash April $11$) that the ratio
$h(\ell=0/\ell=1)=13.95$.  This discrepancy suggests that the numerically
simulated source has higher coherence than that found in the experimental
data. Anyway, we notice that in both epochs the ratio between the two
$\ell=0$ and $\ell=1$ modes of the experimental data are quite stable.

The difference between the theoretical and experimental results can be
attributed to the concomitance of two main effects: 
(1) the experimental source can be more thermalised/scattered than what
is assumed in the theoretical model, suggesting that the emission from
the source has less coherence;
(2) if this loss of coherence depends on the time of separation between
two consecutive acquisitions, our results could have been improved by
reducing $\Delta{t}$ that now is on the order of one day.  Of course,
the acquisition of several snapshots and the reduction $\Delta{t}$
would give more information about the natural coherence of the source
emission mechanism.

What is important in the estimation of the rotation parameter $a$ is not
the coherence of the source \citep{Shvedov:05} but that the asymmetry
between the ${\ell=1}$ and ${\ell=-1}$ modes remain stable. This
property has been observed both in the numerical simulations and with
the experimental results.  The ratio between the ${\ell=0}$ and the
peaks ${\ell=\pm1}$ are not crucial to determine $a$.  This suggests
that the method we used is a promising way to determine the twist
of electromagnetic waves due to the rotation of the BH directly from
the experimental data.

In fact, a better result could be obtained with a fast imaging technique,
involving all EHT radio telescopes, to build up a series of consecutive
snapshots. These images would be characterized to have a short coherence
time due to atmospheric phase fluctuations that can occur on timescales
of tens of seconds.  This technique could be easily implemented in future
EHT observations, with the experimental acquisition of the azimuthal
spatial phase profile of the source, centered on the BH shadow, obtained
with interferometric techniques.  An antenna and data acquisition system
designed for direct measurement of OAM would open entirely new avenues
in astronomy and space sciences and technology in general.

\subsection{Calculating the error in the rotation parameter $a$}
We calculate the total error in the determination of the rotation
parameter $a$ by applying the error propagation theory. To estimate
this error one has to estimate first the error in the spatial phase
distribution on the image plane of an asymptotic observer from the TIE
equations and from the error in the estimation of the intensity $I$.
The $I$ error in each of the EHT images of M87 is obtained
through the convolution of different intensity patterns by using the
NMF (negative matrix factorization) reconstruction method to reduce
the error in agreement with the chi-squared ($\chi^2$) test from EHT
team. This procedure ensures a high-quality image construction process,
with an estimated error in the pattern of $5\%$. The uncertainties are
quantified in units of brightness temperature, with an error confidence
interval of $95\%$ \citep{EHT4}.  Thus, all the images provided by
the EHT collaboration satisfy a $5\%$ maximum error chi-squared test
that is the maximum error we assume for EHT data $| \Delta a |_{EHT}$
in the determination of the intensity spatial distribution.  In summary,
all the images provided by the EHT scientific documentation are presented
and discussed in Refs.~\citep{EHT1,EHT4,EHT5,EHT6}.

As we adopted the numerical approach to solve the TIE equations,
the error is calculated numerically.  In fact, a complete analytic
approach to obtain an exact formulation of the error distribution
needs, of course, the analytic solution to the system of equations
\ref{eq:dI/dz_dP/dz}. This seems not to be possible but for special cases
where the intensity distribution patterns are circular, as discussed
in Ref.~\citep{Ruelas&al:JO:2018}.  The error in the spatial phase
distribution $P(x,y)$ is mainly due to the experimental errors present
in the EHT data.  This error affects the OAM and the spiral spectrum and is
determined numerically by varying, in the input of the TIE routines, the
intensity distribution $I$ with its error interval obtained from public EHT
data. From this we obtain the error in the the spiral spectra,
as reported in the figures~\ref{fig:supp1}-\ref{fig:supp3}. Then
we obtain the error in the estimation of the asymmetry parameter $q$.

The error in the estimation of the parameter $a$ depends on the numerical
error of the TIE reconstruction method, $|\Delta a_{TIE}|$, and from the
error deriving from the final EHT data products, $|\Delta a_{EHT}|$.
The latter term is obtained numerically by  varying $q$ in its error
interval with the polynomial relationship found between the asymmetry
parameter $q$ and the rotation parameter $a$.

We conservatively assume that the final uncertainty of $a$ is given by the
sum of the absolute values of the two errors, the first from the numerical
errors introduced by the routines used to calculate the OAM spectrum
($\sim10^{-12}$) and the polynomial interpolation $|\Delta a_{TIE}|
\sim 10^{-7}$, which can be considered negligible, and the second, which
is due to the error derived from the analysis of the experimental data
provided by EHT and then processed by our numerical routines.

\begin{figure}
\centering
 \includegraphics*[width=1.1\linewidth]{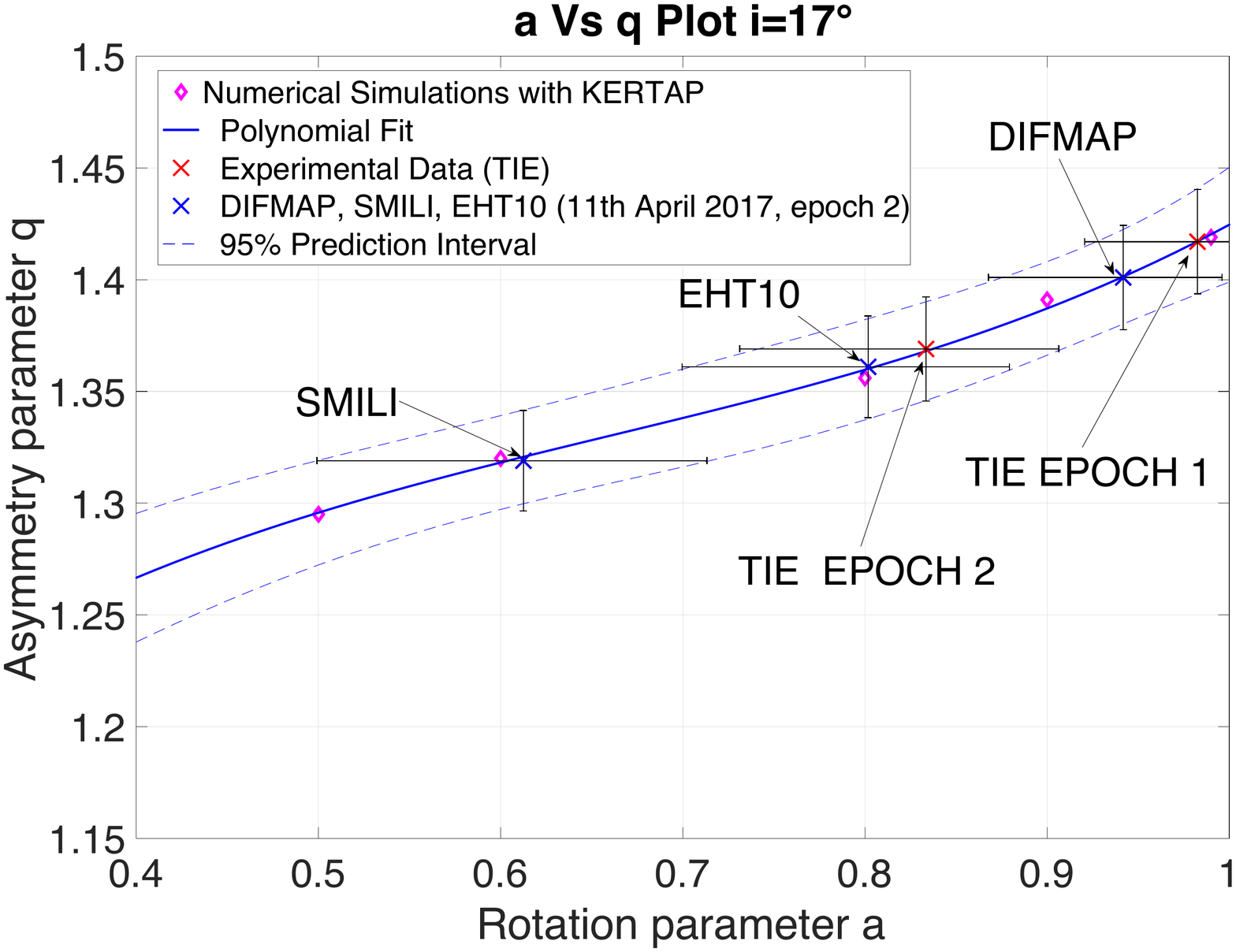}
  \caption{Plots of the rotation parameter $a$ \emph{vs}.\ the asymmetry
   parameter $q$ for ${i=17^\circ}$.  The values of the rotation parameter
   $a$ obtained from the TIE method, and from the visibility amplitude
   and phase maps from the observations on 11th April 2017 obtained by
   EHT with the image processing reduction methods (DIFMAP, SMILI and
   EHT), are in good agreement. They indicate the presence of a disk with
   inclination $i=17^\circ$ (or equivalent to $i=163^\circ$) with $\sim
   95\%$ confidence interval.
  }
\label{fig:supp4}
\end{figure}

\begin{figure}
\centering
 \includegraphics*[width=1.1\linewidth]{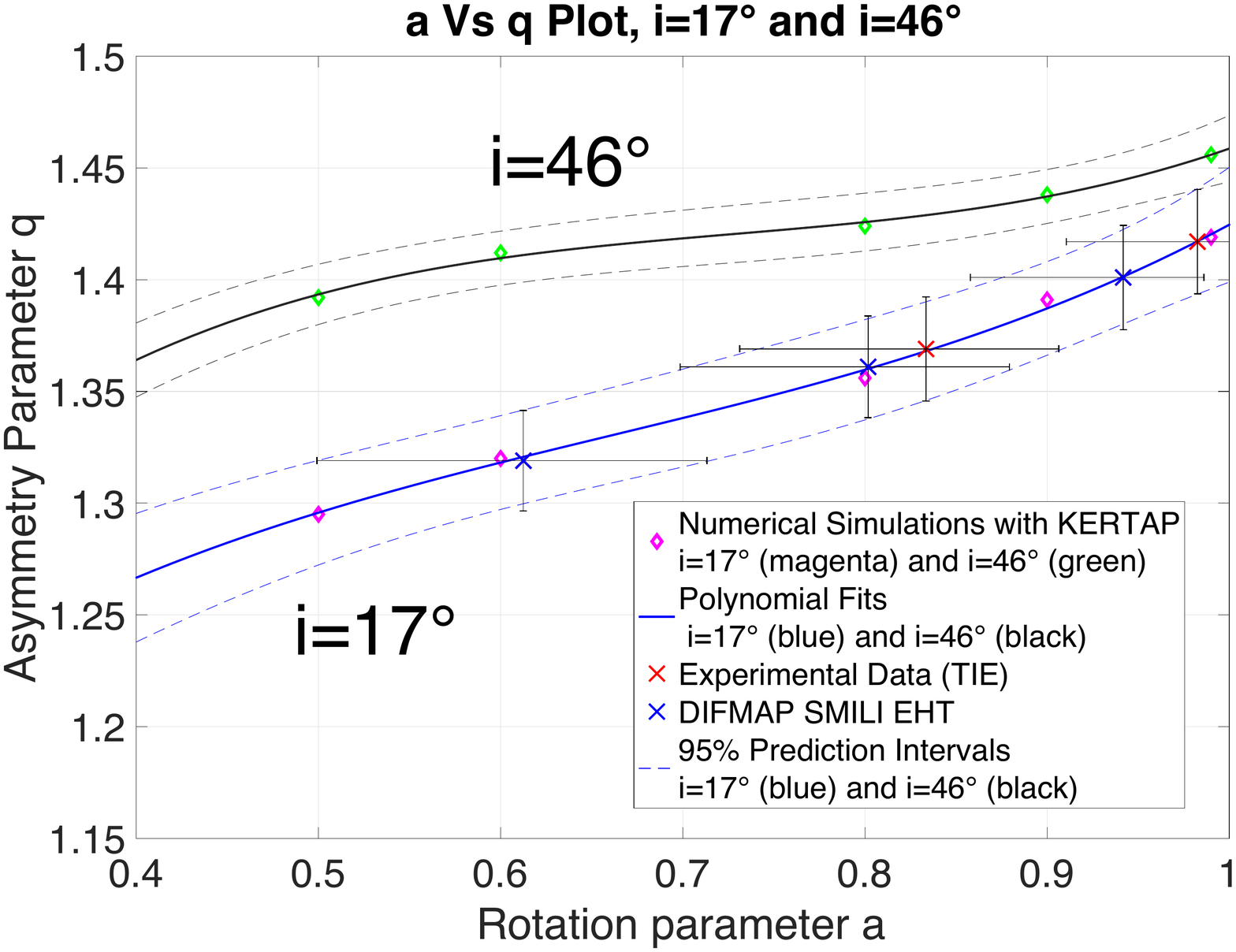}
 \caption{Plots of the interpolations of the rotation parameter $a$
  \emph{vs}.\ the asymmetry parameter $q$ for ${i=17^\circ}$
  and ${i=46^\circ}$ with their $\sim95\%$ confidence intervals.
  The experimental data, presemted in Fig.~\ref{fig:supp4}, favour the
  conclusion of an inclination of ${i=17^\circ}$. Overlapping between the
  two curves and respective $95\%$ confidence intervals are not observed.
 }
\label{fig:supp5}
\end{figure}

As already mentioned, the other term $|\Delta a_{EHT}|$ is calculated
numerically.  To improve the estimate of the parameter $a$ and lower the
associated error, we reduced the maximum error $\sim 5\%$ given in the
data released by EHT in the following way: for each point $P(x,y)$ in any
of the data images of the M87*, we extract the value of the brightness
temperature in $P$ (which is equivalent to the intensity of the field $I$)
and those of the adjacent points in a neighborhood of a chosen set of $N$
pixels, $P(x\pm n,y\pm n)$, where ${n\in[1,\ldots,N]}$.  For the sake of
brevity, we call these intensities $I_j$. Then, we calculate the average
at $P(x,y)$ and the variance
\begin{subequations}
\begin{align}
\overline
 I_{x,y}
&=\frac 1N  \sum_j I_j
\\
 \sigma_{\overline I_{x,y}}^2
&= \frac{\sum_j \sigma_j^2 + 2 \sum_i \sum_{k<j} \mathrm{cov}(X_j, X_k)}{N^2} 
\end{align}
\end{subequations}
If we assume that the $I_j$'s are independent and identically distributed
(iid) random variables, then the respective covariances are zero and we
obtain ${\sigma_{\overline I_{x,y}}^2=\frac{\sum_j \sigma_j^2 }{N^2}}$.
This procedure has been adopted in the numerical routines that are used
to calculate the OAM spectrum and thus the asymmetry parameter $q$ also
for the the spatial phase distribution plots obtained with the TIE method.
After several tests we found that a good compromise both in computational
time and in numerical accuracy is to use $N=3$.  In the calculations of
the OAM spectrum and thus of the asymmetry parameter $q$, from the data
provided by the EHT, we find the maximum error of $q$ being on the order
of $3.4\%$.

From the values of $q$ and their associated error intervals obtained so
far, we then calculate numerically $|\Delta a_{EHT}|$, as reported in
Figure~\ref{fig:supp4}. To calculate the indetermination in the rotation
parameter, we have estimated that the routine used to determine the
intensity at the point $P(x,y)$ through the TIE introduces a maximum
error due to the variable precision arithmetic method of the numerical
routines in Matlab on the order of $\sim10^{-12}$.  The other source
of numerical noise is due to the numerical simulations of the Kerr
spacetime and Einstein ring that are based on the freely available
software package KERTAP \citep{Chen&al:APJS:2015}.  KERTAP characterizes
the propagation of light in Kerr spacetime with very high precision
with a deviation on the order of $10^{-7}$. This contribution that
we call $\left|\Delta{a}\right|_\text{TIE}$ is almost negligible when
compared with the error for $a$ which is introduced by the uncertainties
of the asymmetry parameter $q$, being on the order of $\sim 3.46\%$
and calculated by inserting in the numerical interpolation that relates
the two parameters $a$ and $q$, the values of the edges of the error
bars of each of the two values of $q$ in epoch~1 and epoch~2.  Applying
elementary error theory to the two values of $q$ considered as values of
two independent acquisitions with their errorbars, one easily recovers
$\left|\Delta{a}\right|_\text{EHT}=0.046$.

From the error analysis, the total contribution to the error of $a$ is
composed by these two terms, the one introduced by EHT experimental
errors in the determination of $q$ and $a$ and those due to the small
numerical noise caused by the calculation of the TIE equation and of the
OAM spectrum.  Summing the two contributions, and being the numerical
noise almost negligible, we obtain the total maximum error on the
determination of the rotation parameter dominated by the contribution
introduced by the experimental data,
\begin{equation}
 \left|\Delta{a}\right|_\text{tot}
 =\left|\Delta{a}\right|_\text{TIE}
 +\left|\Delta{a}\right|_\text{EHT} = 0.046
\end{equation}
hence yielding a first estimate of the rotation parameter.  We found that
${a=0.904\pm0.046}$, from the averaged asymmetry in the spiral spectrum
of ${\overline{q}=1.393\pm0.024}$ measured in epoch~1 and epoch~2 as
described in the main text.  We adopt the conservatively approximated
value ${a=0.90\pm0.10}$, as reported in the main text, with an error
with $95\%$ confidence interval as shown in Fig.~\ref{fig:supp4}.

As reported in Table~1 of the main text, different values of the
inclination parameter $i$ yield different families of asymmetry
parameters $q$ that do not overlap, as shown in Fig.~\ref{fig:supp5}.
In this figure we plot the two curves obtained by the interpolation of
the values of $a$ with respect to $q$ for the two values ${i=17^\circ}$
and $i=46^\circ$,  with their respective $95\%$ confidence intervals
and  in the determination of $a$ from $q$ no degeneracy or overlapping
between the two curves, is observed.
The plot also show the experimental data that fit well with the
simulations for the inclination ${i=17^\circ}$.  We point out that to
estimate the rotation parameter $a$ with a single measurement of $q$
if the inclination $i$ is unknown, one has to compare the intensity plots
obtained from the observations with the numerical simulations of the
BH surroundings, as was done by the EHT team.  Another way could be to
use a set of different plots of spatial phase distributions that can be
obtained during a single observation session by dephasing the receiving
antennas and obtain a family of different values of $q$.  This procedure
resembles the methods used during the 2011 public experiment of OAM
radio transmission and detection in San Marco, Venice, by moving one
of the antennas of the interferometer to spatially sample the twisted
wave and tune the OAM radio on the twisted channel from the untwisted
one \citep{Tamburini&al:NJP:2012}.  Of course, a better procedure can
be made by introducing electronic delays in each of the antennas in the
interferometer and/or with a posteriori analysis of the available data.

\bibliographystyle{mnras}
\bibliography{EHT,M87}

\bsp	
\label{lastpage}
\end{document}